\pdfoutput=1
\documentclass[preprint,12pt]{elsarticle}

\usepackage[version=4]{mhchem}
\usepackage{graphicx}%
\usepackage{multirow}%
\usepackage{amsmath,amssymb,amsfonts}%
\usepackage{amsthm}%
\usepackage{mathrsfs}%
\usepackage[title]{appendix}%
\usepackage{xcolor}%
\usepackage{textcomp}%
\usepackage{manyfoot}%
\usepackage{booktabs}%
\usepackage{algorithm}%
\usepackage{algorithmicx}%
\usepackage{algpseudocode}%
\usepackage{listings}%
\usepackage{tabularx,booktabs}
\usepackage{bm}
\usepackage{adjustbox}
\usepackage{tikz}
\usetikzlibrary{arrows.meta,positioning,shapes.geometric,calc, fit}
\definecolor{fwdorange}{HTML}{F24E01}
\definecolor{revblue}{HTML}{0F2D8A}
\newcommand{\fignormal}{\fontsize{8}{9.5}\selectfont}
\newcommand{\figsmall}{\fontsize{7}{8.5}\selectfont}
\usepackage{float}
\usepackage{natbib}

\usepackage{xr-hyper}
\usepackage{hyperref}

\usepackage{booktabs}
\usepackage{tabularx}
\usepackage{array}
\usepackage{ragged2e}

\newcolumntype{L}[1]{%
  >{\RaggedRight\arraybackslash}p{#1}}
\newcolumntype{Y}{%
  >{\RaggedRight\arraybackslash}X}

\newcounter{bla}

\journal{Computer Physics Communications}

\newcommand{\pd}[2]{\frac{\partial #1}{\partial #2}}

\newcommand{\od}[2]{\frac{d #1}{d #2}}

\begin{document}

\begin{frontmatter}

\title{diff-soup: GPU-native Differentiable Simulation and Inference for Large-scale Stiff Dynamical Systems
}

\author[a]{Samuel Degnan-Morgenstern}
\author[a,b]{Martin Z. Bazant\corref{author}}

\cortext[author]{Corresponding author.\\\textit{E-mail address:} bazant@mit.edu}

\address[a]{Department of Chemical Engineering, Massachusetts Institute of Technology, 25 Ames St., Cambridge, MA 02139, USA}
\address[b]{Department of Mathematics, Massachusetts Institute of Technology, 182 Memorial Dr., Cambridge, MA 02139, USA}

\begin{abstract}

Learning physical models from spatiotemporal data requires simulating systems of partial differential equations (PDEs) and differentiating through their solution.
Upon spatial discretization, these models give rise to stiff dynamical systems that require repeatedly solving large, sparse linear systems.
Traditional scientific computing frameworks are largely CPU-based and require problem-specific tuning to accelerate, scale, and differentiate.
In this work, we introduce diff-soup, a differentiable, sparse-oriented, unified PDE suite for GPU-native simulation and inference of large, stiff dynamical systems.
Built in JAX, diff-soup provides 1) a fully differentiable design for problems with sparse structure, 2) scaling to large problems through the NVIDIA cuDSS sparse direct solver with model and data multi-GPU parallelism, and 3) end-to-end PDE-constrained optimization, from classical statistical inference to embedded deep learning.
On stiff PDE benchmarks, diff-soup outperforms established CPU- and GPU-based frameworks by one to two orders of magnitude in both forward solves and gradients.
We demonstrate both a classical Bayesian inference workflow, identifying constitutive relationships, and a modern deep learning workflow, training a neural field for the catalyst activity of a three-dimensional packed bed reactor.
We provide diff-soup as an open-source repository to facilitate research in computational science and engineering.

\vspace{\baselineskip}

\noindent \textbf{PROGRAM SUMMARY}

\begin{small}
\noindent
{\em Program Title:} \texttt{diff-soup}: differentiable, sparse-oriented unified PDE suite \\
{\em CPC Library link to program files:} (to be added by Technical Editor) \\
{\em Developer's repository link:} \url{https://github.com/stmorgenstern/diff-soup} \\
{\em Licensing provisions:} BSD 3-Clause \\
{\em Programming language:} Python, C++ \\
{\em Supplementary material:} The program files in the CPC Program Library contain \texttt{diff-soup} 1.0.0 together with the companion repository \texttt{diff\_soup\_paper}, which holds the benchmark suite, manifests, result stores, figure pipeline and worked examples. \\
{\em Nature of problem:}
Computational modeling of time-dependent partial differential equations (PDEs) gives rise to large systems of stiff ordinary differential equations and differential-algebraic equations.
Robust and efficient integration of these systems requires implicit variable step, variable order (VSVO) methods such as BDF and NDF, whose predominant cost is the repeated factorization and solution of large sparse linear systems.
The established stiff integration libraries run these operations on the CPU, and the general-purpose implicit ODE/DAE time-integration libraries of the \texttt{JAX} ecosystem rely on dense linear algebra, so large two- and three-dimensional problems remain out of reach for GPU-native, differentiable simulation.
These challenges have limited the adoption of machine learning in general multiphysics modeling workflows.\\
{\em Solution method:}
We introduce diff-soup, a general-purpose differentiable, sparse-oriented unified PDE suite for GPU-accelerated multiphysics modeling, simulation, and inference.
We provide a user-friendly Recipe API that models complex physical systems in math-like notation on both structured and unstructured domains with the finite difference, finite volume, and finite element methods.
For fast simulation, we implement sparse Jacobian assembly by automatic differentiation with graph coloring, three VSVO stiff integrators written natively in \texttt{JAX}, and a wrapper to the NVIDIA \texttt{cuDSS} sparse direct linear solver on the GPU.
All simulations are fully differentiable to second order, enabling large-scale PDE-constrained optimization.
Using these derivatives, we provide an inference module for PDE-constrained least-squares problems, from classical frequentist and Bayesian inference to modern deep learning approaches.
Both simulation and inference exploit batched solves, data parallelism, and model parallelism to scale to large problems.\\
{\em Additional comments including restrictions and unusual features:}
This article describes \texttt{diff-soup} 1.0.0.
The package runs on Linux x86\_64 with Python 3.12 and \texttt{JAX} 0.11.
GPU execution requires an NVIDIA GPU with a CUDA 12 or CUDA 13 driver and \texttt{cuDSS} 1.0 or newer, which is installed from NVIDIA's Python wheels.
Without a GPU, every solver runs on the CPU through the KLU and dense backends.
Differential-algebraic equations are supported up to index one.
Solves can be distributed across several GPUs and across the nodes of a cluster.
\end{small}
\end{abstract}

\end{frontmatter}

\section{Introduction}
\label{sec:intro}

The numerical simulation of nonlinear partial differential equations (PDEs) is fundamental to science and engineering, providing quantitative predictions of coupled transport, reaction, and thermodynamic processes across multiple length and time scales.
Important examples include numerical weather and climate prediction \citep{bauer_quiet_2015,palmer_stochastic_2019}, turbulent and reacting flows \citep{moin_direct_1998,lu_toward_2009}, microstructure evolution in materials \citep{cahn_free_1958,chen_phase-field_2002,steinbach_phase-field_2009}, electrochemical energy storage \citep{newman_porouselectrode_1975,doyle_modeling_1993,ramadesigan_modeling_2012}, and biological pattern formation \citep{turing_chemical_1952,kondo_reaction-diffusion_2010}.
Recent advances are bridging these physical models with data-driven methods to enable the learning, design, and control of physical systems directly from data \citep{raissi_physics-informed_2019,karniadakis_physics-informed_2021,rackauckas_universal_2021}.
In weather and climate modeling, learned components are now embedded within physical solvers \citep{reichstein_deep_2019,kochkov_neural_2024}, and in fluid mechanics, neural networks are trained through differentiable solvers to correct and accelerate simulations \citep{brunton_machine_2020,um_solver---loop_2020,kochkov_machine_2021}.
Governing equations and constitutive relationships have been inferred directly from spatiotemporal data \citep{rudy_data-driven_2017,wang_variational_2019,zhao_learning_2023}, and PDE-constrained optimization has been extended from inference to control \citep{holl_learning_2020,cohen_differentiable_2026}.
At the same time, high-resolution imaging techniques such as \textit{operando} microscopy are making spatially resolved data increasingly available for high-dimensional PDE models \citep{liu_review_2019,wu_visualizing_2018,yang_operando_2021}.
Exploiting such datasets requires software that can simulate these models at the scale of the experiment and differentiate through the simulation.

Both requirements are difficult to meet for PDEs, because once discretized in space, they fall into the important class of stiff dynamical systems \citep{byrne_stiff_1987}. 
Stiff problems exhibit multiple competing time scales, such that the slowly varying solution of interest is surrounded by rapidly varying nearby solutions that force a numerical integrator to take small time steps \citep{moler_numerical_2004}.
The integration of stiff systems is a well-established field of computational science, with libraries such as SUNDIALS \citep{hindmarsh_sundials_2005,gardner_enabling_2022}, PETSc \citep{abhyankar_petscts_2018}, \texttt{DifferentialEquations.jl} \citep{rackauckas_differentialequationsjl_2017}, and the MATLAB ODE suite \citep{shampine_matlab_1997} providing specialized implicit integrators.
To both solve and take gradients, these methods require Jacobians which are either manually derived or computed through potentially-inaccurate finite-difference rules, expensive dense automatic differentiation, or performant sparse automatic differentiation.
Gradients additionally require complex, bespoke rules for parameter sensitivities that do not easily compose with stiff solvers.

A central component of these algorithms is the repeated solution of large linear systems.
For PDEs, the local structure of the differential operators gives rise to a discrete sparse representation, which is exploited either by sparse direct solvers, such as KLU \citep{davis_algorithm_2010}, SuperLU \citep{li_overview_2005}, and UMFPACK \citep{davis_algorithm_2004}, or by iterative Newton-Krylov methods like GMRES \citep{saad_gmres_1986,knoll_jacobian-free_2004}.
Sparse direct solvers are robust across a wide range of problems but have traditionally been run on the CPU, with GPU-accelerated factorizations emerging only recently \citep{sao_distributed_2014,ghysels_high_2022}.
Newton-Krylov methods, in contrast, require only Jacobian-vector products and are naturally implemented on GPUs, but their fast and stable convergence depends on a problem-specific preconditioner.

Existing GPU efforts fall into three groups: hybrid codes that keep the integrator on the CPU and offload Krylov kernels, batched solvers for many small systems, and JAX frameworks that target specific physics or rely on dense linear algebra.
SUNDIALS and PETSc support hybrid execution, in which the time integrator remains on the host while large systems are solved with Newton-Krylov methods on the GPU \citep{balos_enabling_2021,mills_toward_2021}.
\texttt{DiffEqGPU.jl} \citep{utkarsh_automated_2024} and \texttt{jaxdae} \citep{li_jaxdae_2026} efficiently batch many small problems on the GPU.
Differentiable PDE solvers such as \texttt{JAX-CFD} \citep{kochkov_machine_2021}, \texttt{JAX-Fluids} \citep{bezgin_jax-fluids_2023}, and \texttt{JAX-FEM} \citep{xue_jax-fem_2023} compose GPU-accelerated simulation with machine learning for specific physics or discretizations, while the general-purpose implicit integrators of \texttt{diffrax} \citep{kidger_neural_2021} rely on dense linear algebra by default.
What remains missing is robust GPU parallelism within a single large, stiff solve, as needed for two- and three-dimensional PDEs, without relying on a problem-specific preconditioner.
To our knowledge, no existing package in the Python/JAX ecosystem natively composes GPU sparse direct solvers, variable step, variable order (VSVO) time integration, and sparse automatic differentiation for the differentiable simulation of large dynamical systems.

The recent release of \texttt{cuDSS} by NVIDIA \citep{nvidia_corporation_cudss_2024}, a sparse direct solver that performs the numerical factorization and the triangular solves on the GPU, removes the major bottleneck in the robust integration of large-scale PDEs.
In this work, we present \texttt{diff-soup}, a GPU-native package for the simulation and inference of large, stiff PDE systems, built around the \texttt{cuDSS} linear solver and sparse automatic differentiation in \texttt{JAX}.
Traditionally, differential equation software has been developed for small model problems and extended to large multiphysics systems afterwards, whereas we take the opposite approach and design every component around the sparse structure of large discretized PDEs.
\texttt{diff-soup} provides the standard numerical tools of a large-scale stiff solver, including VSVO integrators for stiff ordinary differential equations (ODEs) and index-1 differential-algebraic equations (DAEs) with consistent initialization, event handling, and sparse automatic differentiation, together with finite difference, finite volume, and finite element discretizations on grids and meshes.
Beyond these standard tools, \texttt{diff-soup} is distinguished from existing stiff solvers by
\begin{enumerate}
    \item a GPU-native and fully differentiable design that is specialized to problems with sparse structure,
    \item accelerator-specific techniques for reaching large problems, namely reduced precision, batched solves, and multi-GPU parallelism, and
    \item end-to-end workflows for PDE-constrained optimization, with second-order optimizers, multiple shooting, and statistical inference, that compose with \texttt{JAX} libraries such as \texttt{equinox} \citep{kidger_equinox_2021}, \texttt{optax} \citep{deepmind_deepmind_2020}, and \texttt{blackjax} \citep{cabezas_blackjax_2024} for embedded deep learning models and uncertainty quantification.
\end{enumerate}

The remainder of this paper is organized as follows.
Section~\ref{sec:background} reviews the computational background on VSVO time integration, sparse methods, and differentiable simulation.
Section~\ref{sec:diffsoupoverview} describes the modeling, simulation, and inference modules of \texttt{diff-soup}.
In Section~\ref{sec:benchmarks}, we benchmark forward solves and gradients against SUNDIALS, PETSc, \texttt{DifferentialEquations.jl}, MATLAB, and \texttt{diffrax}, and we examine reduced precision and multi-GPU execution.
In Section~\ref{sec:apppdeopt}, we demonstrate two end-to-end applications, the Bayesian inference of the constitutive relationships of a phase-separating material from image data and the training of a neural field embedded in a three-dimensional packed bed reactor model.

\section{Mathematical Background}
\label{sec:background}

In this work, we consider time-dependent systems that take the general form of a differential-algebraic equation
\begin{equation}
    0 = F(y,y', \theta,z(t),t),
    \label{eq:governing-dae}
\end{equation}
where $y \in \mathbb{R}^{n}$ is the state, $t$ is time, $\theta$ is the
parameter vector, and $z(t)$ is the control state that carries externally
prescribed inputs.
These systems are solved on the interval $[t_0,t_f]$ and the initial state is prescribed as $y(t=t_0)=y_0$.
For many systems, this generally reduces to mass-matrix form
\begin{equation}
    M(y,t)\cdot y'= f(y,\theta,z(t),t),
    \label{eq:governing}
\end{equation}
where $M(y,t)$ is the (non-)singular mass matrix.
We partition the state variable into the local differential states, \(y_{d}\in \mathbb{R}^{n_{d}}\), which carry explicit time derivatives in the governing form, local algebraic states, \(y_{a}\in \mathbb{R}^{n_{a}}\), which are specified through algebraic constraints with a sparse structure, and global algebraic variables, \(y_{g}\in \mathbb{R}^{n_{g}}\), \(n_{g}\), which are treated as dense and typically arise from global constraints on the state variables.
Since $n_g \ll n_a + n_d$, this split lets us handle the global variables separately (Sec. \ref{subsec:sparsity}).
In this section, we discuss the relevant background on the three main computational aspects of \texttt{diff-soup}: robust methods for numerical integration of stiff equations, exploiting sparse structure to accelerate computations, and computing sensitivities of differential algebraic equations for PDE-constrained optimization.

\subsection{Variable Step, Variable Order Algorithms}
\label{subsec:vsvo}

In this work, we consider stiff systems with a defined sparse structure, as is typical for the discretization of partial differential equations using the method of lines.
Partial differential equations are naturally stiff due to the decaying spectrum of the underlying differential operators that translates to a large, resolution-dependent ratio between the minimum and maximum eigenvalues \cite{byrne_stiff_1987}.
Stable integration therefore requires specialized methods, since explicit methods are limited by a CFL time scale and naive implicit methods need a prohibitive number of steps to resolve fast dynamics.\cite{shampine_users_1979,beers_numerical_2006}.
We consider the variable step, variable order (VSVO) class of numerical integration algorithms \cite{gear_numerical_1971,byrne_polyalgorithm_1975} which we separate into two families based on the discretization of the time derivative: backward differentiation formula (BDF) \cite{curtiss_integration_1952} in fixed-leading-coefficient (FLC) form \cite{jackson_alternative_1980} and the numerical differentiation formula (NDF) \cite{klopfenstein_numerical_1971} in quasi-constant step size (QSS) form \cite{shampine_matlab_1997}.
We implement three algorithms in \texttt{diff-soup}: the JAX implicit differential algebraic (\texttt{JIDA}) solver based on the popular \texttt{IDA} implementation in SUNDIALS \cite{hindmarsh_sundials_2005}, the JAX BDF solver (\texttt{JBDF}) based on the \texttt{CVODE\_BDF} implementation in SUNDIALS \cite{hindmarsh_sundials_2005} and \texttt{FBDF} implementation in \texttt{DifferentialEquations.jl} \cite{rackauckas_differentialequationsjl_2017}, and the JAX NDF solver (\texttt{JNDF}) based on the \texttt{ode15s} algorithm \cite{shampine_matlab_1997} and implementation of \texttt{QNDF} in \texttt{DifferentialEquations.jl} \cite{rackauckas_differentialequationsjl_2017}.

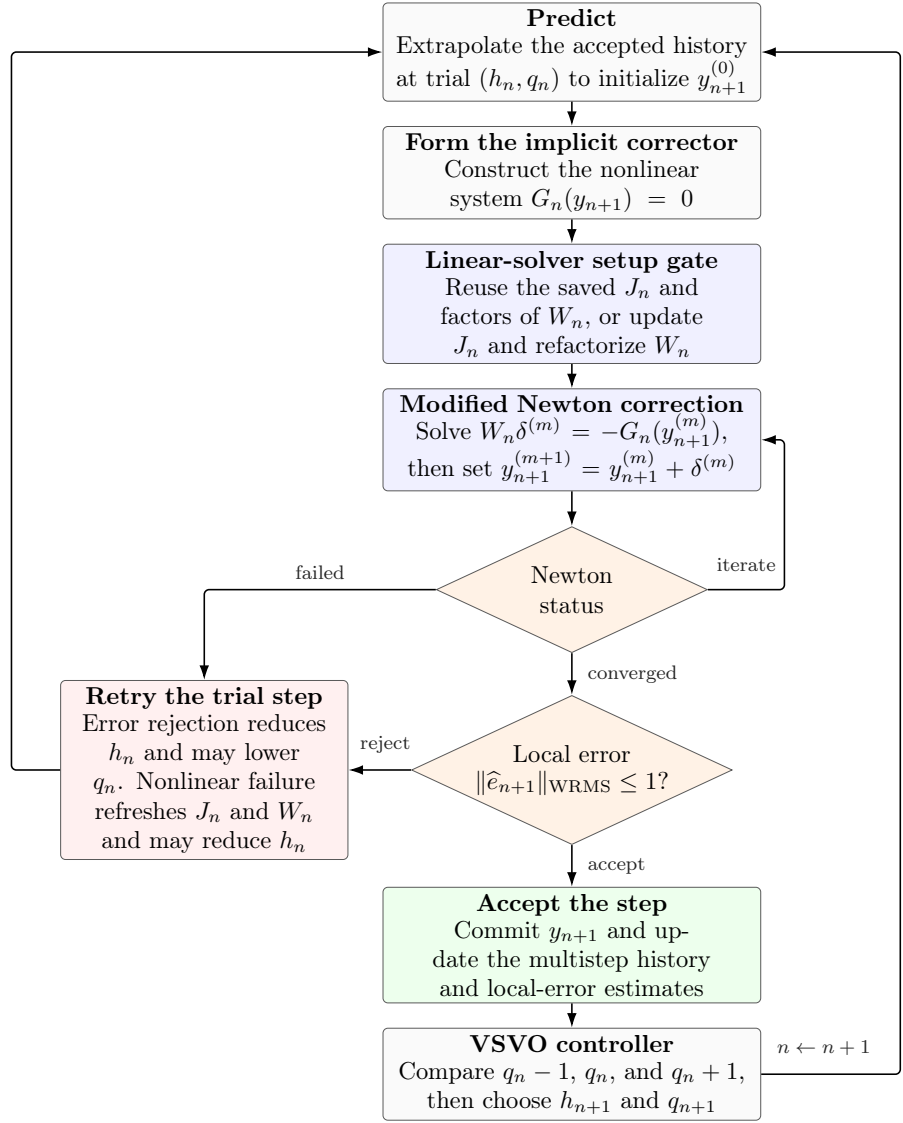
\begin{figure}[H]
\centering

\tikzset{
  vsvo flow/.style={
    -{Latex[length=2.1mm,width=1.4mm]},
    line width=0.65pt,
    rounded corners=2pt
  }
}
\tikzset{
  vsvo process/.style={
    draw=black!65,
    fill=black!2,
    rounded corners=2pt,
    align=center,
    text width=5.15cm,
    minimum height=7.5mm,
    inner xsep=4pt,
    inner ysep=2.7pt
  }
}
\tikzset{
  vsvo solver/.style={
    vsvo process,
    fill=blue!6
  }
}
\tikzset{
  vsvo decision/.style={
    diamond,
    draw=black!65,
    fill=orange!10,
    aspect=2.15,
    align=center,
    text width=2.25cm,
    inner sep=1pt
  }
}
\tikzset{
  vsvo error/.style={
    vsvo decision,
    text width=2.85cm
  }
}
\tikzset{
  vsvo retry/.style={
    vsvo process,
    fill=red!6,
    text width=3.85cm
  }
}
\tikzset{
  vsvo accept/.style={
    vsvo process,
    fill=green!7
  }
}
\tikzset{
  vsvo label/.style={
    font=\scriptsize,
    fill=white,
    text=black!85,
    inner sep=1.1pt
  }
}

\scalebox{0.92}{%
\begin{tikzpicture}[
  font=\footnotesize,
  node distance=3.5mm and 7mm
]

\node[vsvo process] (predict) {
  \textbf{Predict}\\[-1pt]
  Extrapolate the accepted history at trial $(h_n,q_n)$
  to initialize $y_{n+1}^{(0)}$
};

\node[vsvo process, below=of predict] (discretize) {
  \textbf{Form the implicit corrector}\\[-1pt]
  Construct the nonlinear system
  $G_n(y_{n+1})=0$
};

\node[vsvo solver, below=of discretize] (setup) {
  \textbf{Linear-solver setup gate}\\[-1pt]
  Reuse the saved $J_n$ and factors of $W_n$, or update
  $J_n$ and refactorize $W_n$
};

\node[vsvo solver, below=of setup] (newton) {
  \textbf{Modified Newton correction}\\[-1pt]
  Solve
  $W_n\delta^{(m)}=-G_n(y_{n+1}^{(m)})$,
  then set
  $y_{n+1}^{(m+1)}
   =y_{n+1}^{(m)}+\delta^{(m)}$
};

\node[vsvo decision, below=5mm of newton] (status) {
  Newton\\
  status
};

\node[vsvo error, below=6mm of status] (error) {
  Local error\\[-1pt]
  $\|\widehat e_{n+1}\|_{\mathrm{WRMS}}\le 1$?
};

\node[vsvo accept, below=6mm of error] (acceptstep) {
  \textbf{Accept the step}\\[-1pt]
  Commit $y_{n+1}$ and update the multistep history
  and local-error estimates
};

\node[vsvo process, below=of acceptstep] (controller) {
  \textbf{VSVO controller}\\[-1pt]
  Compare $q_n-1$, $q_n$, and $q_n+1$, then choose
  $h_{n+1}$ and $q_{n+1}$
};

\node[vsvo retry, left=9mm of error] (retry) {
  \textbf{Retry the trial step}\\[-1pt]
  Error rejection reduces $h_n$ and may lower $q_n$.
  Nonlinear failure refreshes $J_n$ and $W_n$ and may reduce $h_n$
};

\draw[vsvo flow]
  (predict) -- (discretize);

\draw[vsvo flow]
  (discretize) -- (setup);

\draw[vsvo flow]
  (setup) -- (newton);

\draw[vsvo flow]
  (newton) -- (status);

\draw[vsvo flow]
  (status.south)
  --
  node[
    pos=0.50,
    right=1.8mm,
    vsvo label
  ] {converged}
  (error.north);

\draw[vsvo flow]
  (status.east)
  -- ++(11mm,0)
  node[
    pos=0.50,
    above=2.2mm,
    vsvo label
  ] {iterate}
  |- (newton.east);

\draw[vsvo flow]
  (status.west)
  -- (status.west -| retry.north)
  node[
    midway,
    above=1mm,
    vsvo label
  ] {failed}
  -- (retry.north);

\draw[vsvo flow]
  (error.west)
  --
  node[
    pos=0.41,
    above=2mm,
    vsvo label
  ] {reject}
  (retry.east);

\draw[vsvo flow]
  (error.south)
  --
  node[
    pos=0.50,
    right=1.8mm,
    vsvo label
  ] {accept}
  (acceptstep.north);

\draw[vsvo flow]
  (acceptstep) -- (controller);

\draw[vsvo flow]
  (retry.west)
  -- ++(-7mm,0)
  |- (predict.west);

\draw[vsvo flow]
  (controller.east)
  -- ++(20mm,0)
  node[
    pos=0.45,
    above=2.2mm,
    vsvo label
  ] {$n\leftarrow n+1$}
  |- (predict.east);

\end{tikzpicture}
}

\caption{
Control flow of one trial step in a variable-step, variable-order
implicit multistep integrator.
The predictor extrapolates the accepted history at $(h_n,q_n)$, and a
modified Newton iteration solves the corrector equation using reused or
refreshed sparse LU factors of $W_n$.
Blue boxes mark the linear-solver work that dominates the cost for large systems.
A step attempt can fail from the nonlinear solve if the Newton increments diverge or their contraction rate predicts
that the corrector cannot reach tolerance within the iteration budget.
Additionally, the step can be rejected if the corrector converges, but the weighted local-error estimate exceeds tolerance.
Before retrying the same step, nonlinear failure refreshes $J_n$ and $W_n$ and may reduce $h_n$, while error rejection reduces $h_n$ and may lower $q_n$.
Acceptance commits the history and advances to the next step at $(h_{n+1},q_{n+1})$.
}
\label{fig:vsvo-flow}
\end{figure}

All three algorithms follow a three-stage structure:
Eq.~\eqref{eq:governing-dae} or \eqref{eq:governing} is discretized in time with an implicit multistep formula, the resulting nonlinear system is solved with a modified Newton iteration, and the step size and order are adapted to control the local error.
Figure~\ref{fig:vsvo-flow} shows the control flow of one trial step.
A $k$-step formula approximates the time derivative at $t_{n+1}$ from the current and $k$ stored states,
\begin{equation}
    y'_{n+1} \approx \frac{1}{h_n \beta_0} \sum_{j=0}^{k} \alpha_j\, y_{n+1-j},
    \label{eq:multistep}
\end{equation}
where $h_n$ is the current step size and $\alpha_j$, $\beta_0$ are order-dependent coefficients normalized such that $\alpha_0 = 1$.
The NDF family folds a scaled predictor--corrector difference $(h_n \beta_0)^{-1}\kappa \left(\sum^{k}_{j=1} \frac{1}{j} \right)\bigl(y_{n+1} - y_{n+1}^{(0)}\bigr)$ and its history tangents into $\alpha_j$ and $\beta_0$, where the scalar parameter $\kappa$ trades a modest reduction in the stability angle at higher orders for a smaller truncation-error constant \cite{klopfenstein_numerical_1971,shampine_matlab_1997}.
Substituting Eq.~\eqref{eq:multistep} into Eq.~\eqref{eq:governing} produces a nonlinear algebraic system $G_n(y_{n+1})=0$ whose root advances the state by one step.
Writing $\gamma_n = h_n \beta_0$, each modified Newton iterate solves the linearized system
\begin{equation}
    W_n\,\delta^{(m)} = -G_n\bigl(y_{n+1}^{(m)}\bigr),
    \qquad
    W_n = \tfrac{1}{\gamma_n}\, M - J_n,
    \label{eq:newton}
\end{equation}
and updates $y_{n+1}^{(m+1)} = y_{n+1}^{(m)} + \delta^{(m)}$.
The overall scaling of the corrector equation is purely conventional, since applying Newton's method to $\gamma_n G_n = 0$ produces identical iterates with the rescaled matrix $\gamma_n W_n = M - \gamma_n J_n$, so implementations are free to factorize whichever form of this pencil is most convenient \cite{shampine_implementation_1980}.
For the fully implicit form of Eq.~\eqref{eq:governing-dae}, the same iteration applies with the generalized matrix $W_n = \partial F/\partial y + \tfrac{1}{\gamma_n}\,\partial F/\partial y'$ \cite{petzold_description_1983,brenan_numerical_1996}.
The iteration is warm-started from a predictor extrapolated through the stored history, which is typically accurate enough that a few corrections suffice.

For large systems, the dominant computational bottlenecks are the formation of $J_n$ and the subsequent factorization of the pencil matrix $W_n$ during the linear solve at each iteration of Eq.~\eqref{eq:newton} \citep{byrne_stiff_1987,shampine_implementation_1980}.
The modified Newton iteration freezes both quantities as long as possible, even though reusing the old Jacobian reduces the convergence rate of Newton's method from quadratic to linear \citep{shampine_implementation_1980}.
As such, the development of reliable heuristics for efficient reuse of $J$ across both nonlinear solver iterations and time steps has been an intense area of research and hinges on the interplay between caching rules and the step size and order controllers, which we describe in turn \citep{byrne_stiff_1987,hindmarsh_sundials_2005}.

For the caching rules, all three solvers guard the nonlinear iteration with a two-tier setup gate that re-evaluates the cached Jacobian only when it has aged past a step budget or when a convergence failure has occurred \citep{hindmarsh_sundials_2005}.
Otherwise, when the ratio of the current $\gamma_n$ to the value frozen into the factors of $W_n$ has drifted outside a deadband, the mass-matrix engines rebuild the pencil $\tfrac{1}{\gamma_n}M - J_n$ from the saved $J_n$ without re-differentiation \citep{hindmarsh_sundials_2005}.
\texttt{JIDA} extends the same reuse to the fully implicit form, recomposing $W_n$ from the saved $\partial F/\partial y$ while re-evaluating only $\partial F/\partial y'$, whereas the reference DAE integrator refreshes both partial derivatives at every setup \citep{brenan_numerical_1996,hindmarsh_sundials_2005}.
A convergence-rate memory extrapolates the contraction of the iteration and declares divergence after two iterations rather than exhausting the iteration budget, and failed steps escalate through a ladder that refreshes the Jacobian and shrinks the step \citep{shampine_matlab_1997,hindmarsh_sundials_2005}.
This cost hierarchy shapes the control layer that follows, because any change of the step size or the order moves $\gamma_n$ and invalidates the saved factors, so the controllers are built to change both quantities only as often as accuracy demands.

For the local error control, each solver estimates the local truncation error $\widehat e_{n+1}$ of the converged step from the difference between the predictor and the corrected solution, scaled by an order-dependent constant, and measures it in the weighted root-mean-square (WRMS) norm
\begin{equation}
    \mathrm{err}
    = \Biggl(\frac{1}{n}\sum_{i=1}^{n}
      \Bigl(\frac{\widehat e_{n+1,i}}{w_i}\Bigr)^{2}\Biggr)^{1/2},
    \qquad
    w_i = \mathrm{atol} + \mathrm{rtol}\,\lvert y_i \rvert ,
    \label{eq:wrms}
\end{equation}
where the weight $w_i$ blends the absolute tolerance $\mathrm{atol}$ and the relative tolerance $\mathrm{rtol}$ so that $\mathrm{err} \le 1$ accepts the step as accurate to the requested tolerances \citep{shampine_implementation_1980,hindmarsh_sundials_2005}.
Because the stored history spans the adjacent orders, the same difference tables furnish companion error estimates at orders $q-1$ and $q+1$ at negligible additional cost, and these three estimates drive both the step size and the order selection \citep{gear_numerical_1971,byrne_stiff_1987}.

Each controller converts the error estimate at order $q$ into a proposed growth factor for the step size, $\eta = h_{n+1}/h_n$, through a safety-biased power law of the form $\eta \propto \mathrm{err}^{-1/(q+1)}$, and passes the proposal through a deadband around unity that holds the step size, and thereby the saved factors of $W_n$, unless the predicted change is large enough to justify a refactorization \citep{shampine_implementation_1980}.
The companion estimates at the adjacent orders enter the same way, each candidate order competing through its own estimate.
The solvers differ in three respects, summarized in Table~\ref{tab:vsvo-controllers}: the width and asymmetry of the deadband, the statistic that decides the order, and the handling of startup and failures.
\texttt{JBDF} and \texttt{JIDA} follow their SUNDIALS references \citep{hindmarsh_sundials_2005,brenan_numerical_1996}, while \texttt{JNDF} follows the adaptive scheme of \citet{shampine_matlab_1997} adopted by \texttt{QNDF} \citep{rackauckas_differentialequationsjl_2017}.

\begin{table}[H]
\centering
\footnotesize
\caption{
Comparison of step-size and order control in the three solvers.
After each accepted step, all three form
$\eta \propto \mathrm{err}^{-1/(q+1)}$ and suppress changes in $h$
that do not justify refactorizing $W_n$.
}
\label{tab:vsvo-controllers}

\begingroup
\setlength{\tabcolsep}{3.5pt}
\renewcommand{\arraystretch}{1.04}

\begin{tabularx}{\textwidth}{
  @{}
  L{0.18\textwidth}
  Y
  Y
  Y
  @{}
}
\toprule
&
\multicolumn{1}{c}{\texttt{JBDF}}
&
\multicolumn{1}{c}{\texttt{JIDA}}
&
\multicolumn{1}{c}{\texttt{JNDF}}
\\
\midrule

Deadband
&
Hold for $\eta < 1.5$ and grow otherwise.
An accepted step never reduces $h$.
&
Hold for $1 < \eta < 2$.
Growth is restricted to doubling and reductions to factors in
$[0.5,\,0.9]$.
&
Hold within a constant-step window lasting through $q+2$ consecutive
constant steps.
A steady-state window suppresses mild reductions, but an accepted step
may reduce $h$.
\\
\addlinespace[0.25em]

Order statistic
&
Largest candidate growth factor among $q-1$, $q$, and $q+1$.
&
Smallest scaled truncation-error estimate among $q-1$, $q$, and $q+1$.
&
Largest proposed step size among $q-1$, $q$, and $q+1$.
\\
\addlinespace[0.25em]

Order cadence
&
Only after a wait counter expires.
The wait is $q+1$ steps after an order change and at least two otherwise.
&
Every accepted step.
Order increases require constant-$(h,q)$ history.
&
Every accepted step.
\\
\addlinespace[0.25em]

Startup
&
No separate regime.
The first-step growth cap is large.
&
After the first accepted step, increase the order and double $h$ until
the order stops increasing or a step fails.
&
No separate regime.
\\
\addlinespace[0.25em]

Failure recovery
&
Progressively reduce $h$, possibly lower the order, and restart at order
1 if necessary.
Suppress further changes on the first accepted recovery step.
&
The first rejection reduces $h$ and may lower the order, the second
applies a fixed reduction, and the third forces order 1.
Convergence failures reduce $h$ and refresh $J$.
&
The second consecutive rejection halves $h$, the third lowers the order,
and a fourth rejection at order 1 restarts the history.
\\

\bottomrule
\end{tabularx}

\endgroup
\end{table}

Beneath the controllers, the QSS NDF and FLC BDF families differ in the response of the multistep coefficients in Eq.~\eqref{eq:multistep} to a realized change in the step-size history.
The QSS implementation retains the constant-step coefficients at every step, which makes the formula and its error estimates exact only on a uniform step grid, so it prefers long runs at a fixed step size, defers order changes until the history has run uniformly for several steps, and answers a realized change by rescaling the entire backward-difference history onto the new grid \citep{shampine_matlab_1997}.
The fully variable-coefficient alternative of EPISODE and VODE recomputes every coefficient from the realized nonuniform history, which gives the strongest stability on rapidly varying steps but lets every change in the history perturb the iteration matrix, sacrificing the reuse of the factorization \citep{byrne_polyalgorithm_1975,brown_vode_1989}.
The FLC implementation compromises by pinning the leading coefficient at its constant-step value while the trailing coefficients track the realized history, so that the iteration matrix depends on the step history only through $\gamma_n$ and the stability loss relative to the fully variable-coefficient form is modest \citep{jackson_alternative_1980}.
\texttt{JNDF} follows the QSS convention with the backward-difference table of \citet{shampine_matlab_1997}, while \texttt{JBDF} follows the FLC convention with the Nordsieck history of \citet{hindmarsh_sundials_2005} and \texttt{JIDA} with the divided-difference history of \citet{brenan_numerical_1996}.
We refer the reader to the standard references for complete descriptions of the formulas, controllers, and failure heuristics \citep{gear_numerical_1971,hairer_solving_1996,brenan_numerical_1996,shampine_matlab_1997,hindmarsh_sundials_2005,byrne_stiff_1987}.

\subsection{Exploiting Sparsity}
\label{subsec:sparsity}
As established above, the two dominant computational bottlenecks of the VSVO integration are the formation of the Jacobian $J_n$ and the factorization of the pencil matrix $W_n$ within the modified Newton iteration \citep{byrne_stiff_1987,shampine_implementation_1980}.
Conventional discretization methods for PDEs, such as the finite difference method (FDM), the finite volume method (FVM), and the finite element method (FEM), represent local differential operators in a sparse form such that each component of $y$ couples to a bounded set of neighbors, so the Jacobian $J_n = \partial f/\partial y$ has $O(n)$ nonzero entries in a pattern that is fixed by the discretization \textit{a priori}.
Exploiting this property, we store $J_n$ and $W_n$ in sparse data structures that drastically reduce the memory footprint.
In this section, we describe how \texttt{diff-soup} uses this structure to speed up the Jacobian computation, pencil matrix factorization, and efficiently handle dense, global variables.

Traditionally, differential equation libraries offer three routes to the Jacobian: analytical Jacobians supplied by hand, as in SUNDIALS and PETSc \citep{hindmarsh_sundials_2005,balay_efficient_1997}, which is cumbersome and error-prone for all but the simplest right-hand sides; finite difference approximations, whose truncation and round-off errors degrade the convergence of the Newton iteration \citep{hindmarsh_sundials_2005}; and dense automatic differentiation, which is exact but requires $n$ tangent or cotangent evaluations therefore scaling as $O(n^2)$ \citep{griewank_evaluating_2008}.
More recently, sparse automatic differentiation has matured in the Julia ecosystem, which pairs automatic sparsity detection with graph coloring to compress the Jacobian into a small number of directional derivatives and deploys the approach throughout its differential equation solvers \citep{rackauckas_differentialequationsjl_2017,dalle_common_2026}.

\texttt{diff-soup} builds on this template with the automatic differentiation machinery of \texttt{JAX} \citep{bradbury_jax_2018}.
We assemble the pattern symbolically by composing the declared patterns of the operators that constitute the right-hand side and we compress the Jacobian following the Curtis-Powell-Reid (CPR) approach with a greedy, largest-first coloring of its column intersection graph \citep{curtis_estimation_1974,coleman_efficient_1998,gebremedhin_what_2005}.
The colors define a seed matrix $V \in \{0,1\}^{n \times n_c}$, whose $k$-th column sums the unit vectors of the columns assigned to color $k$, and every nonzero entry of $J$ can be read off from the compressed product
\begin{equation}
    B = J V \in \mathbb{R}^{n \times n_c},
    \label{eq:colored-jacobian}
\end{equation}
because structurally orthogonal columns deposit their entries into disjoint rows of $B$.
For the bounded stencils of FDM and FVM discretizations, $n_c$ is set by the stencil width independent of the resolution, so $n_c \ll n$, and we compute all $n_c$ tangent propagations in parallel with the \texttt{jax.vmap} operator, reducing the cost of the Jacobian from $O(n^2)$ to $O(n)$.
The same batching applies to FEM kernels, where \texttt{jax.vmap} evaluates the element-local residuals across all elements, and the local kernels are differentiated and scattered to assemble the sparse Jacobian in $O(n)$.

Because the global variables couple broadly to the state, they typically produce dense rows and columns that give $J$ an arrowhead structure.
Bidirectional coloring methods can compress such patterns by combining forward and reverse passes \citep{coleman_efficient_1998,gebremedhin_what_2005}, but under the unidirectional column coloring we employ, a single dense row makes every pair of columns share a row, so each column needs its own color.
We therefore exclude the global variables and equations from the sparsity pattern and compute the two dense coupling blocks with local kernels: the $n_g$ dense rows $\partial f_g/\partial y$ are evaluated in reverse mode with one \texttt{jax.jacrev} cotangent per global equation, and the $n_g$ dense columns $\partial f/\partial y_g$ in forward mode with one \texttt{jax.jacfwd} tangent per global variable.

Using this approach, for most PDE problems the remaining computational bottleneck is the factorization of the pencil matrix $W_n$ during the linear solve of Eq.~\eqref{eq:newton}.
Among the classical factorizations, we consider only the LU decomposition, as the pencil matrix $W_n$ is nonsymmetric for general right-hand sides, which precludes the Cholesky factorization, while the QR factorization roughly doubles the operation count and produces denser factors, which reserves it for ill-conditioned least-squares problems \citep{davis_direct_2006}.
An LU solver proceeds in three phases: an analysis phase that determines the pivoting strategy and the structure of the factors, a factorization phase that computes the lower and upper triangular factors $L$ and $U$, and a solve phase that applies them through forward and backward substitution \citep{beers_numerical_2006}.
For a dense matrix, the LU factorization costs $O(n^3)$ while each subsequent solve costs only $O(n^2)$, so the algorithm concentrates the expense in the analysis and factorization to make the solves fast.
This front-loaded cost profile is well matched to the VSVO structure, because the setup gate freezes the pencil across Newton iterations and time steps, so a single factorization is reused for many cheap triangular solves \citep{shampine_implementation_1980,hindmarsh_sundials_2005}.
Nevertheless, the cubic cost of factorization is an immense barrier for large systems.
Much research has focused on linear solvers that exploit sparsity to reduce this cubic cost, and for the purpose of our discussion, we limit ourselves to two families: iterative Newton-Krylov methods and sparse direct solvers \citep{saad_iterative_2003,davis_direct_2006}.

Newton-Krylov methods solve the linearized system in a reduced Krylov subspace $\mathcal{K}_m = \operatorname{span}\{v,\, W_nv,\, W_n^{2}v, \ldots, W_n^{m-1}v\}$ and therefore require only matrix-vector products with $W_n$, and hence only Jacobian-vector products, which makes them attractive for large systems \citep{knoll_jacobian-free_2004}.
The workhorse algorithm implemented in most libraries is the generalized minimal residual (GMRES) method, which builds an orthonormal basis of $\mathcal{K}_m$ and selects the iterate that minimizes the residual norm over the subspace, imposing no symmetry requirements on $W$ \citep{saad_gmres_1986}.
However, the convergence of the Krylov iteration deteriorates on the ill-conditioned spectra that stiff discretizations produce, and much work has been devoted to preconditioners that transform the system before the iteration \citep{saad_iterative_2003}.
The best preconditioners are often problem specific, since their goal is to cluster the eigenvalues of the preconditioned system near unity, and the two most common generic choices are the incomplete LU (ILU) factorization and algebraic multigrid (AMG), for which we refer the reader to \citet{saad_iterative_2003} and \citet{ruge_algebraic_1987} for further details.

In contrast to the iterative methods, sparse direct solvers operate on the sparse representation of $W$ itself and require no problem-specific tuning.
Sparse analogs of the LU factorization, such as the KLU algorithm, retain the analysis, factorization, and solve structure while confining the computation to the nonzero matrix entries \citep{davis_algorithm_2010}.
In general, the factors of a sparse matrix are generally much denser than the matrix itself due to fill in during elimination, so the analysis phase must compute a fill-reducing ordering and pay careful attention to the interaction between ordering, pivoting, and sparsity \citep{davis_direct_2006}.
For matrices arising from two- and three-dimensional meshes under nested dissection orderings, the factorization costs $O(n^{3/2})$ and $O(n^{2})$ respectively in the total number of unknowns $n$, a substantial reduction from the dense $O(n^3)$ although still superlinear \citep{george_nested_1973,davis_direct_2006}.
Sparse elimination is inherently serial and branching heavy, so these methods have traditionally struggled to exploit CPU and GPU parallelism.
\texttt{SuperLU\_MT} achieved a CPU-parallel factorization by scheduling the independent subtrees of the elimination tree across threads and pipelining the remaining panel updates, and we refer the reader to \citet{demmel_asynchronous_1999} for the details of the algorithm.
Practical GPU implementations have emerged only recently, most notably NVIDIA’s cuDSS, a GPU-accelerated sparse direct solver.
We have developed \texttt{diff-soup} around \texttt{cuDSS}, mapping its three phases onto the caching structure of the integrator: the symbolic analysis is performed once per sparsity pattern, which the discretization fixes \textit{a priori}, the numeric factorization is repeated only when the setup gate refreshes $W_n$, and the cached factors serve every Newton iterate in between through cheap triangular solves.
This solver has also allowed us to explore GPU-specific strategies for accelerating the computation, such as working in reduced precision, batching independent solves in parallel, and employing multi-GPU execution, as we discuss in Section~\ref{sec:benchmarks}.

Finally, the partitioning of the state into local and global variables allows us to accelerate the linear solves through a Schur complement.
We partition the Newton system of Eq.~\eqref{eq:newton} conformally with the state as
\begin{equation}
    \begin{pmatrix} A & B \\ C & D \end{pmatrix}
    \begin{pmatrix} \delta_l \\ \delta_g \end{pmatrix}
    = -\begin{pmatrix} G_l \\ G_g \end{pmatrix},
    \label{eq:schur-partition}
\end{equation}
where $l$ collects the local differential and algebraic states and $g$ the global variables.
Through the pencil $W_n = \tfrac{1}{\gamma_n}M - J_n$, each block inherits the structure of the corresponding Jacobian block: $A$ carries the sparse local coupling of $\partial f_l/\partial y_l$, $D$ is the small dense $n_g \times n_g$ global block of $\partial f_g/\partial y_g$, and $B$ and $C$ carry the dense coupling columns $\partial f_l/\partial y_g$ and rows $\partial f_g/\partial y_l$.
Eliminating $\delta_l$ from the first block row and substituting into the second reduces the system to the Schur complement $S$ of $A$,
\begin{equation}
    S \,\delta_g = -G_g + C A^{-1} G_l,
    \qquad
    S = D - C A^{-1} B,
    \label{eq:schur-solve}
\end{equation}
after which the local increment is recovered as \(\delta _{l}=-A^{-1}G_{l}-(A^{-1}B)\delta _{g}\).
In practice, we factorize $A$ only when the setup gate refreshes $W_n$ and cache the $n_g$ sparse solves forming $A^{-1}B$ alongside the factors, so each Newton iterate costs one sparse solve, one small dense solve, and the back-substitution.
Since $n_g \ll n_a + n_d$, the Schur system adds negligible cost, and the dense rows and columns never enter the sparse factorization, preserving the local sparsity and the fill-reducing ordering of $A$ \citep{davis_algorithm_2010}.

\subsection{Differentiable Simulation and PDE-constrained Optimization}
\label{subsec:backpdeopt}
Recent progress in scientific machine learning has concentrated on differentiable simulation for PDE-constrained applications in learning, inference, design, control, and uncertainty quantification \citep{rackauckas_universal_2021,sapienza_differentiable_2024,zhao_image_2021,holl_learning_2020,cohen_differentiable_2026}.
The PDE-constrained optimization problem takes the form
\begin{align}
    &\min_{\theta}\; \mathcal{L}\bigl(y(t,\theta),\theta\bigr), \nonumber\\
    &\;\text{s.t.} \quad M(y,t)\cdot y' = f(y,\theta,z(t),t),
    \label{eq:pde-opt}
\end{align}
where $\mathcal{L}$ is a scalar cost function and $\theta \in \mathbb{R}^{n_\theta}$ collects the parameters of interest.
Conventional first-order optimization requires the total derivative of the cost with respect to the parameters,
\begin{equation}
    \frac{d}{d\theta}\,\mathcal{L}\bigl(y(t,\theta),\theta\bigr)
    = \pd{\mathcal{L}}{y}\,\pd{y}{\theta} + \pd{\mathcal{L}}{\theta},
    \label{eq:loss-gradient}
\end{equation}
where the central quantity is the sensitivity matrix $\partial y/\partial \theta \in \mathbb{R}^{n \times n_\theta}$, the derivative of the solution with respect to the parameters.

To compute the sensitivity matrix, one must differentiate Eq.~\eqref{eq:governing} with respect to the parameters using either the continuous or discrete sensitivity paradigm.
Optimize-then-discretize derives continuous auxiliary equations for the sensitivities and integrates them numerically alongside the state, as in the CVODES and IDAS integrators of SUNDIALS \citep{serban_cvodes_2008,hindmarsh_sundials_2005}
Discretize-then-optimize differentiates the discrete computational graph of the solver program and can be implemented in three ways: hand-derived discrete adjoints, as in PETSc's \texttt{TSAdjoint} \citep{abhyankar_petscts_2018,zhang_petsc_2022} and FATODE \citep{zhang_fatode_2014};  symbolic derivation of the adjoint from the recorded variational forms of a finite element solve, as in \texttt{dolfin-adjoint} \citep{farrell_automated_2013}; and automatic differentiation (AD) of the solver's computational graph, as in \texttt{diffrax} and this work \citep{kidger_neural_2021,bradbury_jax_2018}.
The Julia ecosystem exposes both routes behind a common sensitivity interface \citep{rackauckas_differentialequationsjl_2017,rackauckas_universal_2021}.
Both the continuous and discrete paradigms can be computed in forward mode, where one tangent per parameter is propagated forward, or reverse (adjoint) mode, where one cotangent per scalar cost traverses backwards.
This gives four methods as described in detail in this section and compared in Table \ref{tab:sensitivity-methods}: forward sensitivity analysis (FSA) and adjoint sensitivity analysis (ASA) for continuous sensitivities \citep{cao_adjoint_2003,hindmarsh_sundials_2005}, and forward and reverse AD for discrete sensitivities, as implemented in  \texttt{diff-soup} \cite{kidger_neural_2021}. 

FSA differentiates the governing equations with respect to each parameter.
For a state-independent mass matrix, differentiating Eq.~\eqref{eq:governing} with respect to $\theta_j$ yields the tangent system
\begin{equation}
    M\,s'_j = \pd{f}{y}\, s_j + \pd{f}{\theta_j},
    \label{eq:fsa}
\end{equation}
where $s_j = \partial y/\partial \theta_j \in \mathbb{R}^{n}$ is the $j$-th column of the sensitivity matrix, and the fully implicit form of Eq.~\eqref{eq:governing-dae} differentiates the same way into $\partial F/\partial y\, s_j + \partial F/\partial y'\, s'_j + \partial F/\partial \theta_j = 0$.
The tangent systems are linear and share the pencil matrix $W$ with the forward solve, so the sensitivities advance alongside the state through cheap solves with the cached factors \citep{hindmarsh_sundials_2005}.
The sensitivity matrix is also useful beyond the gradient, as for least-squares cost functions the product $\left(\partial y/\partial\theta\right)^{T}\left(\partial y/\partial\theta\right)$ is the canonical Gauss-Newton approximation to the Hessian.
Because the integrator carries one tangent system per parameter, the cost of FSA scales as $O(n_\theta\, T(n))$, where $T(n)$ denotes the cost of one forward VSVO solve under the sparse methods above.

\begin{figure}[H]
\centering
\tikzset{
  stepblock/.style={draw, thick, rounded corners, minimum width=1.35cm,
    minimum height=0.9cm, align=center, font=\fignormal},
  zoomstep/.style={stepblock, fill=red!12},
  gapdots/.style={font=\fignormal},
  bigbox/.style={draw, thick, rounded corners, align=center,
    inner sep=5pt, text width=3.6cm, font=\fignormal},
  rootbox/.style={draw, thick, rounded corners, align=center,
    inner sep=5pt, text width=3.0cm, minimum height=3.2cm, font=\fignormal},
  rulebox/.style={draw, thick, rounded corners, align=center,
    inner sep=5pt, text width=3.8cm, minimum height=3.2cm, font=\fignormal},
  zoombox/.style={draw, densely dotted, rounded corners, inner sep=8pt},
  lvl/.style={font=\fignormal\bfseries},
  tangentflow/.style={-Stealth, thick, fwdorange},
  cotangentflow/.style={-Stealth, thick, revblue},
  zoomline/.style={densely dashed, gray},
}
\begin{adjustbox}{max width=\textwidth}
\begin{tikzpicture}
  \node[stepblock] (s1) {step $1$};
  \node[gapdots, right=0.9cm of s1] (d1) {$\cdots$};
  \node[zoomstep, right=0.9cm of d1] (s3) {step $n{+}1$};
  \node[gapdots, right=0.9cm of s3] (d2) {$\cdots$};
  \node[stepblock, right=0.9cm of d2] (s4) {step $N$};
  \node[stepblock, right=1.0cm of s4] (cost) {$\mathcal{L}$};
  \coordinate (in) at ($(s1.west)+(-1.0cm,0)$);

  \draw[tangentflow] ([yshift=2.5mm]in) -- node[above, font=\figsmall, text=black]
    {$y_0,\ \textcolor{fwdorange}{\dot y_0}$} ([yshift=2.5mm]s1.west);
  \draw[tangentflow] ([yshift=2.5mm]s1.east) -- node[above, font=\figsmall, text=black]
    {$y_1,\ \textcolor{fwdorange}{\dot y_1}$} ([yshift=2.5mm]d1.west);
  \draw[tangentflow] ([yshift=2.5mm]d1.east) -- node[above, font=\figsmall, text=black]
    {$y_n,\ \textcolor{fwdorange}{\dot y_n}$} ([yshift=2.5mm]s3.west);
  \draw[tangentflow] ([yshift=2.5mm]s3.east) -- node[above, font=\figsmall, text=black, pos=0.75]
    {$y_{n+1},\ \textcolor{fwdorange}{\dot y_{n+1}}$} ([yshift=2.5mm]d2.west);
  \draw[tangentflow] ([yshift=2.5mm]d2.east) -- ([yshift=2.5mm]s4.west);
  \draw[tangentflow] ([yshift=2.5mm]s4.east) -- node[above, font=\figsmall, text=black]
    {$y_N,\ \textcolor{fwdorange}{\dot y_N}$} ([yshift=2.5mm]cost.west);

  \draw[cotangentflow] ([yshift=-2.5mm]cost.west) -- node[below, font=\figsmall]
    {$\bar y_N,\ \bar\theta_N$} ([yshift=-2.5mm]s4.east);
  \draw[cotangentflow] ([yshift=-2.5mm]s4.west) -- ([yshift=-2.5mm]d2.east);
  \draw[cotangentflow] ([yshift=-2.5mm]d2.west) -- node[below, font=\figsmall, xshift=2mm]
    {$\bar y_{n+1},\ \bar\theta_{n+1}$} ([yshift=-2.5mm]s3.east);
  \draw[cotangentflow] ([yshift=-2.5mm]s3.west) -- node[below, font=\figsmall, xshift=-2mm]
    {$\bar y_n,\ \bar\theta_n$} ([yshift=-2.5mm]d1.east);
  \draw[cotangentflow] ([yshift=-2.5mm]d1.west) -- node[below, font=\figsmall]
    {$\bar y_1,\ \bar\theta_1$} ([yshift=-2.5mm]s1.east);
  \draw[cotangentflow] ([yshift=-2.5mm]s1.west) -- node[below, font=\figsmall]
    {$\bar y_0,\ \bar\theta_0$} ([yshift=-2.5mm]in);

  \node[zoombox, fit=(in)(s1)(cost), inner sep=0.5cm] (bframe) {};

  \node[bigbox] (sim) at ($(bframe.north)+(0,1.8cm)$) {%
    \textbf{differentiable simulator}\\[2pt]
    adaptive VSVO solve of Eq.~\eqref{eq:governing} for $y$};
  \node[stepblock, right=2.1cm of sim] (loss) {$\mathcal{L}$};
  \coordinate (ain) at ($(sim.west)+(-2.3cm,0)$);
  \draw[tangentflow] ([yshift=2.5mm]ain) -- node[above, font=\figsmall, text=black]
    {$y_0,\ \theta,\ \textcolor{fwdorange}{\dot y_0,\ \dot\theta}$} ([yshift=2.5mm]sim.west);
  \draw[cotangentflow] ([yshift=-2.5mm]sim.west) -- node[below, font=\figsmall]
    {$\bar y_0,\ \bar\theta$} ([yshift=-2.5mm]ain);
  \draw[tangentflow] ([yshift=2.5mm]sim.east) -- node[above, font=\figsmall, text=black]
    {$y,\ \textcolor{fwdorange}{\dot y}$} ([yshift=2.5mm]loss.west);
  \draw[cotangentflow] ([yshift=-2.5mm]loss.west) -- node[below, font=\figsmall]
    {$\bar y$} ([yshift=-2.5mm]sim.east);

  \node[rootbox] (root) at ($(bframe.south)+(0,-2.8cm)$) {%
    \textbf{accepted-step root}\\[4pt]
    $G(y_{n+1}) = 0$\\[6pt]
    implicit function\\ theorem:\\[2pt]
    $\dfrac{d y_{n+1}}{d\theta} = -\,W^{-1}\,\od{G}{\theta}$\\[2pt]
    \strut};
  \node[rulebox, draw=fwdorange, left=1.05cm of root] (jvp) {%
    \textbf{JVP rule}\\[4pt]
    $(y_n,\ \textcolor{fwdorange}{\dot y_n},\ \textcolor{fwdorange}{\dot\theta}) \mapsto$\\
    $(y_{n+1},\ \textcolor{fwdorange}{\dot y_{n+1}})$\\[6pt]
    $W\, \dot y_{n+1} = -\pd{G}{y_n}\, \dot y_n - \pd{G}{\theta}\, \dot\theta$\\[6pt]
    \strut\\[3pt]
    \strut};
  \node[rulebox, draw=revblue, right=1.05cm of root] (vjp) {%
    \textbf{VJP rule}\\[4pt]
    $(y_{n+1},\, \textcolor{revblue}{\bar y_{n+1}},\, \textcolor{revblue}{\bar\theta_{n+1}}) \mapsto$\\
    $(\textcolor{revblue}{\bar y_n},\ \textcolor{revblue}{\bar\theta_n})$\\[6pt]
    $W^{T} \lambda = -\bar y_{n+1}$\\[2pt]
    $\bar y_n \mathrel{+}= \pd{G}{y_n}^{T} \lambda$\\[2pt]
    $\bar\theta_n \mathrel{+}= \pd{G}{\theta}^{T} \lambda$};
  \draw[tangentflow] ([xshift=-0.9cm, yshift=2.5mm]root.west) -- node[above, font=\figsmall]
    {$\dot\theta$} ([yshift=2.5mm]root.west);
  \draw[cotangentflow] ([yshift=-2.5mm]root.west) -- node[below, font=\figsmall]
    {$\bar\theta$} ([xshift=-0.9cm, yshift=-2.5mm]root.west);
  \draw[tangentflow] ([yshift=2.5mm]root.east) -- node[above, font=\figsmall]
    {$\dot y$} ([xshift=0.9cm, yshift=2.5mm]root.east);
  \draw[cotangentflow] ([xshift=0.9cm, yshift=-2.5mm]root.east) -- node[below, font=\figsmall]
    {$\bar y$} ([yshift=-2.5mm]root.east);

  \node[zoombox, inner sep=6pt, fit=(jvp)(root)(vjp)] (czoom) {};

  \draw[zoomline] ([xshift=-4mm]sim.south) -- (bframe.north -| sim);
  \draw[zoomline] ([xshift=4mm]sim.south) -- (bframe.north -| sim);
  \draw[zoomline] ([xshift=-4mm]s3.south) -- (czoom.north -| s3);
  \draw[zoomline] ([xshift=4mm]s3.south) -- (czoom.north -| s3);

  \node[lvl] at ($(czoom.west |- sim)+(-0.35cm,0)$) {(a)};
  \node[lvl] at ($(czoom.west |- s1)+(-0.35cm,0)$) {(b)};
  \node[lvl] at ($(czoom.west |- root)+(-0.35cm,0)$) {(c)};
\end{tikzpicture}
\end{adjustbox}
\caption{
Differentiation through the discrete computational graph of the adaptive integration loop.
(a) At the highest level, the simulator is one differentiable map, where the forward solve consumes the initial state $y_0$ and the parameters $\theta$ to produce the trajectory $y$ that enters the cost $\mathcal{L}$, forward mode pushes the seeds $(\dot y_0, \dot\theta)$ through to the solution tangent $\dot y$, and reverse mode pulls the cost gradient $\bar y = \pd{\mathcal{L}}{y}^{T}$ back to $(\bar y_0, \bar\theta)$.
(b) The simulator unrolls into the chain of accepted steps, where each step passes the state forward with its tangent alongside (orange) while the cotangents and the running parameter cotangent flow backward (blue), so $\bar\theta_0$ exits as the assembled gradient.
(c) Inside one accepted step, the implicit function theorem binds both derivative rules at the corrector root $G(y_{n+1}) = 0$: the JVP rule, the pushforward of forward mode, maps the incoming state and tangents into the new state and tangent through one linear solve with the exact pencil $W$, and the VJP rule, the pullback of reverse mode, consumes the replayed state with the incoming cotangents, solves the transposed system for the multiplier $\lambda$, and adds this step's contributions to the running cotangents $\bar y_n$ and $\bar\theta_n$.
The total derivative $\od{G}{\theta}$ holds $y_{n+1}$ fixed and collects both the direct parameter dependence $\pd{G}{\theta}$ and the indirect dependence through the stored history.
The multistep history generalizes the single-lag dependence shown, as in Eq.~\eqref{eq:ift-tangent}.
}
\label{fig:ad-through-solver}
\end{figure}
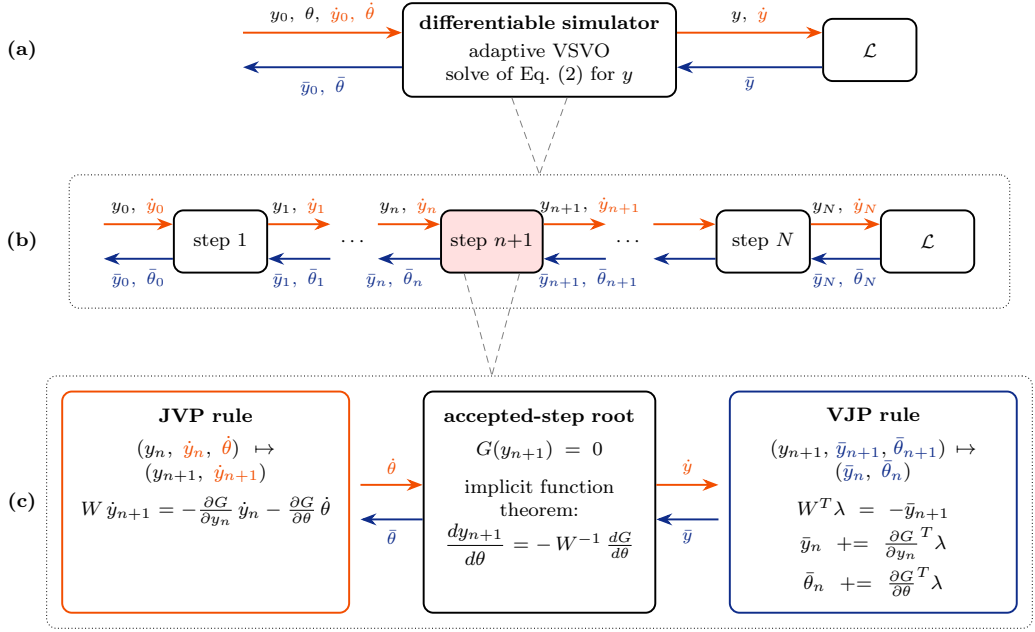

Alternatively, forward-mode AD propagates tangents through the discrete computational graph of the adaptive integration loop, which \texttt{diff-soup} implements as a \texttt{lax.while} loop \citep{bradbury_jax_2018} as shown in Figure~\ref{fig:ad-through-solver}.
The tangent $\dot y_n = \pd{y_n}{\theta}\,\dot\theta \in \mathbb{R}^{n}$ is defined as the directional derivative of the accepted state along a seed direction $\dot\theta \in \mathbb{R}^{n_\theta}$, and forward mode advances one tangent per seed \citep{griewank_evaluating_2008}.
Unrolled, the loop is a chain of accepted steps, and each step is an implicit function of the parameters and of the $k$ stored history states, because the corrector equation $G_n(y_{n+1}) = 0$ inherits the stored states through the multistep formula and the parameters through the right-hand side.
Rather than naively differentiating through the Newton algorithm, we apply the implicit function theorem at the converged root \citep{kidger_neural_2021,blondel_efficient_2022}, which turns the constant mass matrix step dependence into the tangent recurrence
\begin{equation}
    W_n\, \dot y_{n+1}
    = \pd{f}{\theta}\, \dot\theta
    - \tfrac{1}{\gamma_n}\, M \sum_{j=1}^{k} \alpha_j\, \dot y_{n+1-j},
    \label{eq:ift-tangent}
\end{equation}
where $W_n$ is the exact pencil of Eq.~\eqref{eq:newton} evaluated at the root, the first term carries the direct parameter dependence, and the sum carries the stored history tangents into the new step.
To ensure the correctness of the derivative, a fresh pencil matrix $W_n$ and factorization must be used at an extra computational cost rather than the potentially stale factor used by the forward program.
We employ the frozen-controller approximation, such that we apply gradient truncation to the adaptive order and step size control flow.
Instead, tangents differentiate the realized control schedule as a fixed discretization \citep{kidger_neural_2021}.
Because forward mode AD carries one tangent per seed, the cost retains the $O(n_\theta\, T(n))$ scaling of FSA.
To reduce practical cost, we leverage the shared pencil factorization to efficiently batch the tangent computation at each step.

For systems with many parameters, such as training modern deep neural networks, the $O(n_\theta\, T(n))$ scaling of the forward methods renders differentiation practically infeasible.
Fortunately, reverse-mode and adjoint methods compute the gradient of a scalar cost directly, at a cost that no longer multiplies in $n_\theta$.
In the continuous setting, adjoint sensitivity analysis introduces the costate $\lambda(t) \in \mathbb{R}^{n}$ and, for a constant mass matrix and a cost of integral form $\mathcal{L} = \int_{t_0}^{t_f} g(y,\theta,t)\,dt$, integrates the adjoint system backward from $t_f$,
\begin{equation}
    M^{T}\lambda' = -\pd{f}{y}^{T}\lambda - \pd{g}{y}^{T},
    \qquad
    \lambda(t_f) = 0,
    \label{eq:adjoint}
\end{equation}
after which the gradient follows from the quadrature
\begin{equation}
    \frac{d\mathcal{L}}{d\theta}
    = \int_{t_0}^{t_f} \Bigl(\pd{g}{\theta} + \lambda^{T}\pd{f}{\theta}\Bigr)\,dt
    + \lambda(t_0)^{T} M\, \pd{y_0}{\theta},
    \label{eq:adjoint-gradient}
\end{equation}
where $y_0$ carries any parameter dependence of the initial state and terminal costs enter through a nonzero terminal condition.
For a singular mass matrix, the terminal condition must itself be consistently initialized, because the algebraic components of the costate are determined by the adjoint constraints, and we refer the reader to \citet{cao_adjoint_2003} for the derivation in the fully implicit and higher-index settings.

The backward solve of Eq.~\eqref{eq:adjoint} requires the forward trajectory $y(t)$, and the implementations trade memory against recomputation through three strategies: a dense interpolant of the forward solution, checkpointing with segment recomputation, and re-integration of the state backward in time alongside the adjoint \citep{sapienza_differentiable_2024,rackauckas_differentialequationsjl_2017}.
The Julia ecosystem pairs the interpolant with an adaptive Gauss quadrature of the gradient integral in Eq.~\eqref{eq:adjoint-gradient}, which typically converges in far fewer evaluations than the time integration itself \citep{rackauckas_differentialequationsjl_2017,kim_stiff_2021}.
For stiff problems, however, two stability concerns constrain these choices: the backward re-integration of the state, popularized for neural ODEs \citep{chen_neural_2018}, is unstable because the reversed dissipative dynamics amplify small forward errors, and the adjoint system inherits the stiffness of the forward problem, so it must itself be integrated implicitly with the same sparse linear algebra as the forward solve \citep{kim_stiff_2021}.
The number of costates is parameter independent and quadrature accumulates the $n_\theta$ gradient components additively, so the total cost scales as $O(T(n) + n_\theta)$.

In the JAX-native engines, we instead differentiate the adaptive control loop in reverse mode as a discrete adjoint using the checkpointed while loop in \texttt{equinox}, which implements the optimal online checkpointing algorithms of \citet{stumm_new_2010} and \citet{wang_minimal_2009}.
In reverse mode, the tangent solve of Eq.~\eqref{eq:ift-tangent} transposes, and each step propagates the cotangent $\bar y_{n+1} = \od{\mathcal{L}}{y_{n+1}}^{T}$ backward through one linear solve with $W_n^{T}$, traversing the chain of Figure~\ref{fig:ad-through-solver} from the cost to the parameters.
Rather than storing the trajectory, the loop retains $O(\sqrt{n_{\mathrm{steps}}})$ checkpoints of the accepted states by default, where $n_{\mathrm{steps}}$ is the user-prescribed step ceiling, and recomputes the segments between them during the backward sweep, bounding the memory at $O(\sqrt{n_{\mathrm{steps}}}\, n)$.
One backward sweep costs a constant recomputation of the forward solve and accumulates the parameter cotangent $\bar\theta \in \mathbb{R}^{n_\theta}$ additively across the steps, so the gradient scales as $O(T(n) + n_\theta)$.
As in forward mode AD, we employ the frozen-controller assumption, so cotangents flow only through the accepted steps, and we refer the reader to \citet{kidger_neural_2021} and the \texttt{diffrax} documentation for a detailed treatment of these checkpointing and control-flow choices.

We summarize the computational cost, memory footprint, accuracy, and stability concerns of the four sensitivity algorithms in Table~\ref{tab:sensitivity-methods}.
None of the four algorithms returns the exact gradient of Eq.~\eqref{eq:pde-opt}, though the source of the error differs between the two families.
The continuous methods approximate the gradient of the continuous problem to a tolerance, with error governed by the tolerances of the selected forward solver, sensitivty solver, and quadrature method\citep{cao_adjoint_2003,sapienza_differentiable_2024}.
The discrete methods differentiate the frozen discretization selected by the forward solve exactly, to floating point precision, and their gradients deviate from that of the adaptive solution map by a term controlled by the local error tolerance.

\begin{table}[H]
\centering
\footnotesize
\caption{
Comparison of the four sensitivity methods for the PDE-constrained optimization problem of Eq.~\eqref{eq:pde-opt}.
$T(n)$ denotes the cost of one forward VSVO solve under the sparse methods of this section, so no entry incurs a dense cubic factorization.
Costs and memory are reported in the state dimension $n$, the parameter count $n_\theta$, and the user-prescribed step ceiling $n_{\mathrm{steps}}$.
The discrete methods differentiate the realized discretization under frozen step size and order controls.
}
\label{tab:sensitivity-methods}

\begingroup
\setlength{\tabcolsep}{3.5pt}
\renewcommand{\arraystretch}{1.04}

\begin{tabularx}{\textwidth}{
  @{}
  L{0.14\textwidth}
  Y
  Y
  Y
  Y
  @{}
}
\toprule
&
\multicolumn{1}{c}{FSA}
&
\multicolumn{1}{c}{Forward AD}
&
\multicolumn{1}{c}{ASA}
&
\multicolumn{1}{c}{Reverse AD}
\\
\midrule

Cost
&
$O(n_\theta\, T(n))$.
Tangents share the pencil factorization.
&
$O(n_\theta\, T(n))$.
Tangents batched into multi-RHS solves with \texttt{jax.vmap}.
&
$O(T(n) + n_\theta)$.
One backward solve plus quadrature.
&
$O(T(n) + n_\theta)$.
A constant recomputation overhead from checkpointing.
\\
\addlinespace[0.25em]

Memory
&
$O(n_\theta\, n)$ live tangents.
No trajectory storage.
&
$O(n_\theta\, n)$ live tangents.
No trajectory storage.
&
Dense interpolant of $O(n_{\mathrm{steps}}\, n)$, or checkpoint-reduced.
&
$O(\sqrt{n_{\mathrm{steps}}}\, n)$ checkpoints by default; budget prescribable.
\\
\addlinespace[0.25em]

Gradient computed
&
Continuous sensitivities, accurate to the integration tolerances.
&
Exact tangents of the realized discretization under frozen controls.
&
Continuous gradient, accurate to the forward, adjoint, and quadrature tolerances.
&
Exact gradient of the realized discretization under frozen controls.
\\
\addlinespace[0.25em]

Stiffness concerns
&
Inherits the stability of the forward solve.
&
Inherits the stability of the forward solve.
&
The adjoint system is itself stiff, and backward re-integration of the state is unstable.
&
Stable.
Replays stored or recomputed forward states.
\\

\bottomrule
\end{tabularx}

\endgroup
\end{table}
Beyond smooth trajectories, the control state $z(t)$ introduces events, such as protocol switches and state-triggered conditions, that stop the integration, update the state, and restart it from a consistent state.
Because the frozen-controller assumption fixes the time grid, the discrete computational graph alone would miss how the switch time moves with the parameters, so we attach custom JVP and VJP rules to each event.
These rules differentiate the switch time by the implicit function theorem and carry its shift into the restarted trajectory, so tangents and cotangents pass through an event as through any accepted step of Figure~\ref{fig:ad-through-solver}.
At transversal crossings, the resulting derivatives match the classical event sensitivities of \citet{galan_parametric_1999} and \citet{hiskens_trajectory_2000} and their adjoint counterparts \citep{corner_adjoint_2020}
This machinery is required for future applications in control protocol design, where the optimization variables move the events themselves.

Finally, we support the computation of discrete second-order sensitivities, which second-order optimization methods for Eq.~\eqref{eq:pde-opt} consume as Hessian-vector products or as the assembled Hessian.
We implement three second-order compositions through the solver: forward-over-forward differentiation, which propagates second-order tangents along the realized trajectory and assembles the full $n_\theta \times n_\theta$ Hessian at $O(n_\theta^{2}\, T(n))$, reverse-over-forward differentiation, which computes gradients of scalar criteria built from the solution and its forward sensitivities through one reverse sweep over the tangent solve at $O(n_\theta\, T(n))$, and forward-over-reverse differentiation, the canonical strategy for Hessian-vector products \citep{griewank_evaluating_2008}, which differentiates the checkpointed gradient in forward mode and delivers each product at a constant multiple of the gradient cost, $O(T(n) + n_\theta)$, so Newton-type methods extract curvature without ever assembling the Hessian.

\section{diff-soup Software Overview}
\label{sec:diffsoupoverview}
\lstdefinestyle{dspython}{
  language=Python,
  basicstyle=\scriptsize\ttfamily,
  keywordstyle=\bfseries,
  commentstyle=\itshape,
  showstringspaces=false,
  columns=fullflexible,
  keepspaces=true,
  breaklines=false,
  frame=none,
  xleftmargin=1em,
  numbers=none,
}
\floatstyle{plain}
\newfloat{listing}{tbp}{lol}
\floatname{listing}{Listing}

\begin{figure}[H]
\centering
\includegraphics[width=\textwidth]{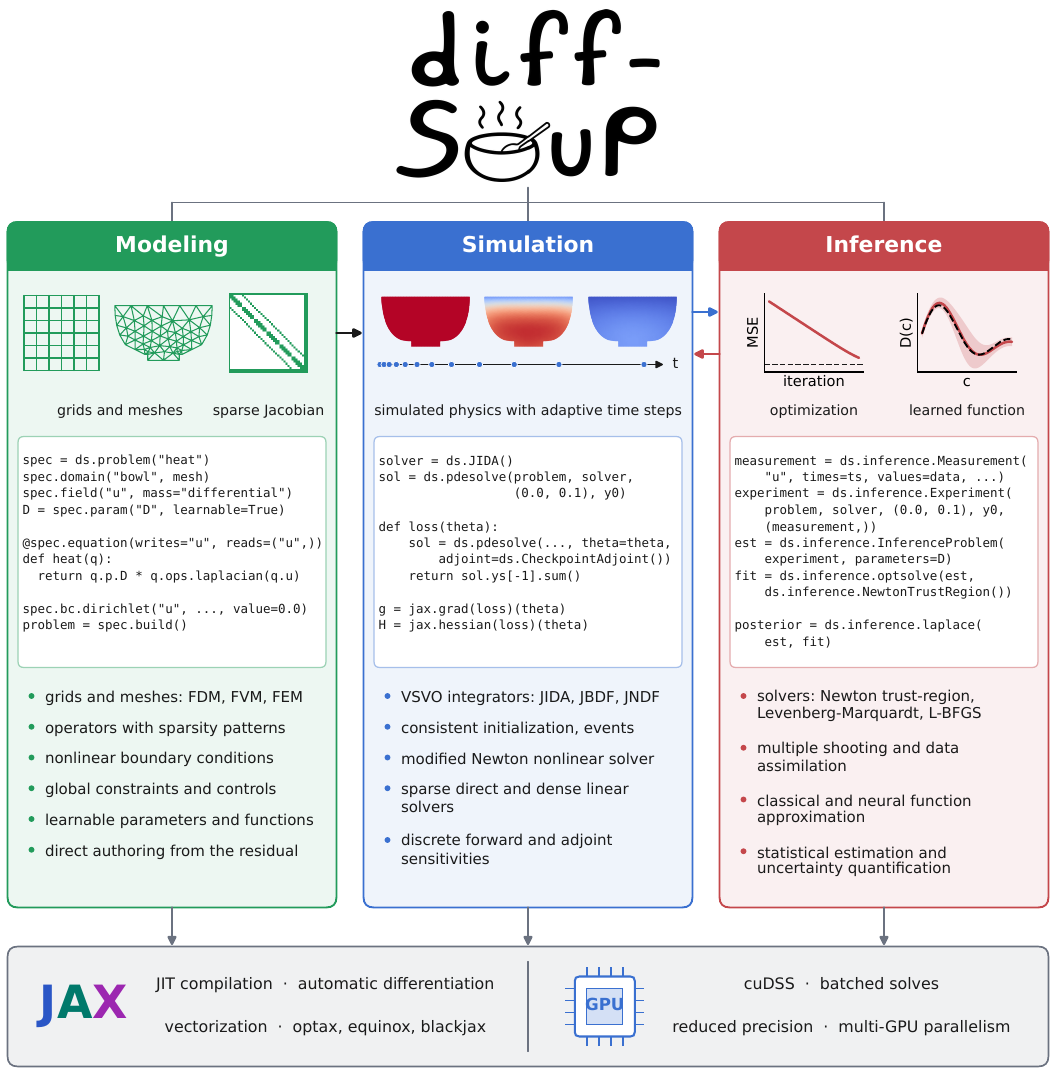}
\caption{Visual overview of the three primary modules of \texttt{diff-soup} for modeling, differentiable simulation, and inference.}
\label{fig:software-overview}
\end{figure}

In this section, we introduce the architecture of \texttt{diff-soup}, specifically the three primary modules for modeling, simulation, and inference as visually outlined in Figure~\ref{fig:software-overview}.
The modeling module provides the tools for declaring a PDE system, its geometry, and its discretization method through \texttt{diff-soup}'s Recipe API to construct a sparse, discrete problem for the time-integrators.
The simulation module integrates this system in time using the stiff VSVO algorithms and provides discrete parameter sensitivities.
Finally, the inference module solves PDE-constrained optimization problems to learn model parameters from data.

\subsection{Modeling}
\label{subsec:modeling}

In this subsection, we describe the modeling module, the primary user entry point in \texttt{diff-soup}, and its lowering to an internal representation (IR) system used for simulations.
We have designed the Recipe API as a user-friendly workflow for specifying physical equations in math-like notation without needing a symbolic algebra system.
For example, consider the heat equation on the unit interval with diffusivity $D$,
\begin{equation}
\begin{aligned}
    \pd{u}{t} &= D\,\nabla^2 u,   &&\quad x \in (0,1),\; t > 0, \\
    u(0,t)    &= u(1,t) = 0,      \\
    u(x,0)    &= x(1-x),          
\end{aligned}
\label{eq:heat-recipe}
\end{equation}
which translates to the following \texttt{diff-soup} Recipe:
\begin{listing}[H]
\begin{lstlisting}[style=dspython]
import numpy as np
import diff_soup as ds
from diff_soup.domains import cartesian_grid_1d

n = 64
spec = ds.problem("heat")                                  # open the builder
spec.domain("rod", cartesian_grid_1d(n, 0.0, 1.0, method="fvm"))  # 64 cells
spec.field("u", mass="differential", scale=1.0, init=lambda x: x * (1.0 - x))
D = spec.param("D", value=0.1, learnable=True, bounds=(0.01, 2.0))  # learnable

@spec.equation(writes="u", reads=("u",))
def heat(q):
    return q.p.D * q.ops.laplacian(q.u)                    # du/dt = D lap(u)

spec.bc.dirichlet("u", node_indices=np.array([0, n - 1]), value=0.0)

problem = spec.build()                                     # lower the recipe
\end{lstlisting}
\caption{
The heat equation of Eq.~\eqref{eq:heat-recipe} declared through the Recipe API.
The unit interval is discretized on 64 finite-volume cells, the diffusivity is declared as a learnable parameter with box bounds, both ends are pinned to zero, and \texttt{build} lowers the recipe into the problem that the simulation module consumes.
}
\label{lst:modeling}
\end{listing}

First, the user specifies the domain on which the physical system will be simulated.
\texttt{diff-soup} provides a flexible domain and discretization system with support for both structured grids through FDM, FVM, and FEM, and unstructured meshes through FEM.
Once a domain is registered, the unknowns are declared on it as fields, scalar- or vector-valued variables that are specified as differential or algebraic variables.
The user then writes the PDE system in strong form as a plain Python closure over an access object $q$, composing the field values, the declared parameters, and the differential operators of the domain.
\texttt{diff-soup} maintains a readily extendable library of common differential operators for science and engineering, with each operator declaring the induced sparsity pattern for composition into the sparse Jacobian discussed in Sec. \ref{subsec:sparsity}.
Finally, boundary conditions are attached to each field, selected by node index on a grid or by named tags on a mesh, with \texttt{diff-soup} providing the standard linear and nonlinear boundary conditions such as Dirichlet, Neumann, Robin, periodic, and flux boundary conditions.

For more complex workflows, we provide additional modeling features for declaring global constraints, externally controlled inputs, and learnable parameters or parameterized functions.
For nonlocal constraints, such as prescribed domain or boundary integrals, the user declares global algebraic variables and a corresponding global constraint to close the system.
Since constraints couple densely, we handle global variables using the Schur complement, as described in Section~\ref{subsec:sparsity}, to avoid compromising local sparsity.
We also provide controls for external forcing protocols in two forms, a direct control for an input variable read inside the equations and a derived control for a quantity the solver assembles.
Furthermore, for differentiable simulation we require the user to mark explicitly which parameters and parameterized functions are learnable, and optionally specify box bounds for the optimizer.

Internally, spec.build() lowers the Recipe to an internal representation that provides the residual, sparse Jacobian, learnable parameters, and controls to the solver.
By default, an equation closure returns the right-hand side of the mass-matrix form of Eq.~\eqref{eq:governing}, and if an equation uses the symbolic time derivative of a field, the fully implicit form of Eq.~\eqref{eq:governing-dae} is used.
For advanced use cases, mass-matrix and implicit problems can be authored directly from the residual and sparsity pattern.

\subsection{Simulation}
\label{subsec:simulation}

In this subsection, we describe the simulation module, which integrates the physical system in time, and exposes a JIT-compiled, differentiable function for the inference module.
A simulation is run by calling \texttt{pdesolve} on the lowered problem and providing a desired solver algorithm, the time interval, and an initial state, as shown in Listing~\ref{lst:simulation}.
\texttt{diff-soup} implements three native VSVO solvers: \texttt{JIDA}, \texttt{JBDF}, and \texttt{JNDF}, as discussed in Section~\ref{subsec:vsvo}.
Additionally, we provide a wrapper to \texttt{diffrax} \cite{kidger_neural_2021}, for verification and non-stiff problems.
For DAEs, the first step in each simulation is consistent initialization, in which the differential variables are held at their initial values and the resulting nonlinear system of algebraic variables and differential slopes is solved to satisfy all algebraic equations and constraints.
Each time step then follows the control flow of Figure~\ref{fig:vsvo-flow}.
The modified Newton corrector solves the sparse linear system with cuDSS on the GPU, KLU on the CPU, or optionally a dense solver, and handles the global variables through the Schur complement of Section~\ref{subsec:sparsity}.
Finally, events interrupt the integration when a user-specified condition is met to terminate the solve or modify the state and controls before continuing.

\begin{listing}[H]
\begin{lstlisting}[style=dspython]
import jax
import numpy as np

y0 = spec.assemble_state({})                        # the initial state
solver = ds.JIDA()                                  # VSVO DAE engine
ts = np.linspace(0.02, 0.1, 5)
sol = ds.pdesolve(problem, solver, (0.0, 0.1), y0, saveat=ds.SaveAt(ts=ts))
sol.ys, sol.ok, sol.stats                           # (5, 64) states, outcome, stats

theta = ds.default_theta(problem)                   # the parameter pytree

def final_state(theta):
    return ds.pdesolve(problem, solver, (0.0, 0.1), y0, theta=theta).ys[-1]

S = jax.jacfwd(final_state)(theta)                  # forward mode (default)

def loss(theta):
    sol = ds.pdesolve(problem, solver, (0.0, 0.1), y0, theta=theta,
                      adjoint=ds.CheckpointAdjoint())  # checkpointed reverse mode
    return sol.ys[-1].sum()

g = jax.grad(loss)(theta)                           # one backward sweep
H = jax.hessian(loss)(theta)                        # forward over reverse
\end{lstlisting}
\caption{
A solve and differentiation of the running heat equation example given in Eq.~\eqref{eq:heat-recipe}.
The JIDA solver integrates the problem of Listing~\ref{lst:modeling} to five output times, forward mode differentiates the final state with respect to the parameters at the default settings, and the checkpointed adjoint enables the reverse-mode gradient and the forward-over-reverse Hessian of a scalar loss.
}
\label{lst:simulation}
\end{listing}

The \texttt{pdesolve} function is compatible with the standard \texttt{JAX} transformations including JIT compilation, automatic differentiation, and batching through \texttt{jax.vmap}.
The compiled solve is cached for each problem and solver configuration, so the time interval, initial conditions, parameters, and controls can change without recompilation.
To differentiate through the solve, we employ the discretize-then-optimize approach of Section~\ref{subsec:backpdeopt} with a frozen controller.
By default, forward mode automatic differentiation propagates one tangent per input direction through the adaptive loop, with batching used when possible.
Reverse mode uses a discrete checkpointed adjoint to compute the gradient of a scalar objective with respect to all parameters in one backward pass within a user-prescribed memory budget.
Second-order sensitivities compose both forward and reverse AD, with forward-over-reverse differentiation the default method for computing Hessian-vector products at a constant multiple of the gradient cost, and full Hessians.
Additionally, events are differentiable, with tangents and cotangents carried through the crossing by the implicit-function rule at the trigger root of Section~\ref{subsec:backpdeopt}.
Many independent solves are batched through \texttt{jax.vmap} into one program that advances every solve together and factorizes the batch of matrices in a single call.
GPU acceleration is further exploited through a reduced precision mode to enable computations in Float64, Float32, or mixed precision mode -- which factorizes in Float32 precision and does iterative refinement with the Float64 precision residual.
Finally, \texttt{diff-soup} can distribute computation across multiple GPUs using both data and model parallelism to either distribute independent solves across GPUs or spread a single sparse factorization across several devices on one node or across the nodes of a cluster, as we quantify for one node in Section~\ref{sec:GPUaccel}.

\subsection{Inference}
\label{subsec:inference}
Next, we describe the inference module of \texttt{diff-soup}, which is designed to solve the PDE-constrained optimization problem of Eq.~\eqref{eq:pde-opt} for learning physics-based models from data.
To estimate the diffusivity value in the model heat equation problem, we employ the workflow shown in Listing~\ref{lst:inference}.

\begin{listing}[H]
\begin{lstlisting}[style=dspython]
# data: the solution at D = 0.3 at the times ts, plus Gaussian noise of scale 1e-3
measurement = ds.inference.Measurement("u", times=ts, values=data,
                                       likelihood=ds.inference.Gaussian(sigma=1e-3))
experiment = ds.inference.Experiment(problem, solver, (0.0, 0.1), y0, (measurement,))
est = ds.inference.InferenceProblem(experiment, parameters=D)   # maximum likelihood
fit = ds.inference.optsolve(est, ds.inference.NewtonTrustRegion())
fit[D], fit.status                                  # fitted D, termination status

posterior = ds.inference.laplace(est, fit)          # Gaussian approximation at the fit
ident = ds.inference.identifiability(est, fit)      # the information spectrum
\end{lstlisting}
\caption{
A fit of the running heat equation example given in Eq.~\eqref{eq:heat-recipe} and its uncertainty quantification.
The observed measurement is declared as synthetic data generated with D = 0.3 at 5 time points with Gaussian noise. 
The corresponding experiment is registered, and the inference problem is posed to recover the maximum likelihood estimate of the diffusivity using a Newton trust-region solver.
Following the inference problem, a variety of diagnostics are used to assess the solution including posterior estimation and a practical identifiability report.
}
\label{lst:inference}
\end{listing}
First, the observed data streams are registered as measurement objects and collected together with the corresponding lowered problem in an experiment object.
Using one or more experiments, the user then defines an inference problem over the parameters of interest and solves it using \texttt{ds.inference.optsolve} with the desired first-order or second-order optimization algorithm.
After solving, the result object provides the optimized parameters, the cost history, a termination status, and diagnostics of the solution quality.

We provide extensive support for posing the inference problem as a statistical estimation problem to determine the maximum likelihood estimate (MLE) under a frequentist approach or the maximum a posteriori (MAP) estimate under a Bayesian approach.
To compute the MLE, a likelihood can be specified, as in Listing~\ref{lst:inference}.
To compute the MAP, in addition to the likelihood, priors can be declared on the parameters.
After the inference problem is solved, a Laplace approximation to the posterior can be derived from the Hessian or Gauss-Newton Hessian approximation around the MAP.
For more sophisticated posterior estimation, the posterior can be sampled using the Markov chain Monte Carlo No-U-Turn sampler (NUTS) in \texttt{blackjax} with preconditioning from the Laplace covariance.
Samples are used to create credible intervals around the fitted parameters.
Additionally, practical identifiability can be assessed from the Fisher information matrix and from the profile likelihood of each parameter, in which that parameter is fixed and all others are refit.

We implement three native optimization algorithms in \texttt{diff-soup}:  a Newton trust-region method using the exact Hessian or its Gauss-Newton approximation, a Levenberg-Marquardt method, and a limited-memory BFGS method.
Additionally, we provide wrappers to the \texttt{optax} and \texttt{optimistix} libraries for first-order training of embedded deep learning models and for further solvers.

We provide a variety of tools and modules in \texttt{diff-soup} for typical inference workflows including data assimilation, batching, and function approximation.
Specifically, we provide pre-built tools for projecting image data onto a mesh, refining the mesh, and estimating initial and boundary conditions from image data.
For batching, multiple shooting is employed to partition large trajectories into shorter windows that are joined 
by continuity constraints.
This both reduces the computational cost of long trajectory simulations and improves the robustness and conditioning of the optimization problem \citep{bock_multiple_1984,cohen_differentiable_2026}.
In addition, we offer a variety of prebuilt modules for approximating functions and operators that commonly occur in PDEs.
Orthogonal polynomials, neural networks, and neural operators are supplied for approximating constitutive relationships, functions, and operators within PDEs.
We provide Karhunen–Loève expansions and coordinate networks for approximating spatial fields both classically and using deep learning.
Additionally, a variety of differentiable, piecewise interpolants such as constant, linear, and cubic polynomials are offered for time-varying signals.

As discussed in Section~\ref{subsec:simulation}, \texttt{diff-soup} employs both data and model parallelism for distributing computation across multiple GPUs.
Under data parallelism, subsets of a larger dataset, either distinct experiments or windows for multiple shooting, are distributed across all GPUs, and the losses, gradients, and curvature are then summed across the devices.
We specialize this for hierarchical inference problems where the dataset consists of many experiments on distinct domains with multiple episodes per domain.
The problem can be distributed into three types of parameters: global parameters shared by all experiments, cluster parameters that are shared by a single domain, and local parameters that are specific to a distinct episode.
This allows us to exploit the arrowhead structure of the resulting joint Hessian and employ a block Newton solver that uses a Schur complement over the shared block.
Under model parallelism, a single simulation whose sparse factorization outgrows the memory resources of one GPU is factored across several devices.

\section{Benchmarks}
\label{sec:benchmarks}
In this section, we benchmark \texttt{diff-soup} against widely used scientific computing frameworks for stiff dynamical systems, including the CPU libraries SUNDIALS \citep{hindmarsh_sundials_2005}, PETSc \citep{abhyankar_petscts_2018}, \texttt{DifferentialEquations.jl} \citep{rackauckas_differentialequationsjl_2017}, and MATLAB \texttt{ode15s} \citep{shampine_matlab_1997}, and the GPU-based \texttt{JAX} library \texttt{diffrax} \citep{kidger_neural_2021}.
We use three problems of increasing complexity: the heat equation, which admits a closed-form solution in both the continuous and the discrete setting, the Brusselator system, which is a model benchmark problem for stiff PDEs \citep{hairer_solving_1996}, and a challenging coupled chemo-mechanical Cahn-Hilliard problem inspired by PFHub benchmark problem 4 \citep{jokisaari_phase_2018}.

\subsection{Comparing time integrators with the heat equation}
\label{subsec:bench-heat}

\begin{figure}[H]
\centering
\includegraphics[width=0.63\linewidth]{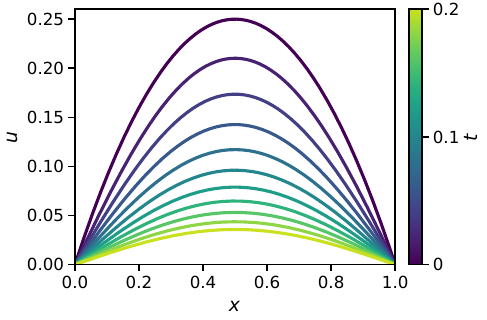}
\caption{
Solution of the heat equation of Eq.~\eqref{eq:heat-bench} at the eleven output times of the benchmark, evenly spaced over $t \in [0, 0.2]$ and colored by time.
}
\label{fig:heat-evolution}
\end{figure}

The heat equation on the unit interval with homogeneous Dirichlet boundaries takes the following form:
\begin{equation}
\begin{aligned}
    \pd{u}{t} &= \frac{\partial^2 u}{\partial x^2}, &&\quad x \in (0,1),\; t \in (0, T], \\
    u(0,t)    &= u(1,t) = 0, \\
    u(x,0)    &= x(1-x),
\end{aligned}
\label{eq:heat-bench}
\end{equation}
which is discretized with second-order central differences on $N$ interior nodes $x_i = i\,\Delta x$ with spacing $\Delta x = 1/(N+1)$ as
\begin{equation}
     u_i' = \frac{u_{i-1} - 2u_i + u_{i+1}}{\Delta x^2},
    \qquad i = 1, \ldots, N,
    \qquad u_0 = u_{N+1} = 0,
\label{eq:heat-discrete}
\end{equation}
and integrated to $T = 0.2$.
Both the continuous problem and the semidiscrete system admit closed-form solutions,
\begin{equation}
\begin{aligned}
    u(x,t) &= \sum_{k=1}^{\infty} \frac{4\,(1 - (-1)^k)}{k^3 \pi^3}\, \sin(k \pi x)\, e^{-k^2 \pi^2 t}, \\
    u_i(t) &= \sum_{k=1}^{N} \hat u_k\, \sin(k \pi x_i)\, e^{-\lambda_k t},
    \qquad
    \lambda_k = \frac{4}{\Delta x^2} \sin^2\!\left(\frac{k \pi \Delta x}{2}\right),
\end{aligned}
\label{eq:heat-solutions}
\end{equation}
where $\hat u_k$ are the discrete sine coefficients of the initial condition and $\lambda_k$ are the eigenvalues of the negative discrete Laplacian.
Figure~\ref{fig:heat-evolution} shows the decay of the solution over the benchmark window. 

We note that the eigenvalues of the discrete Laplacian span $\lambda_1 \approx \pi^2$ to $\lambda_N \approx 4/\Delta x^2$, which causes the stiffness, typically characterized by the ratio of the largest to smallest eigenvalues, to grow as $O(N^2)$ \citep{leveque_finite_2007}.
Furthermore, explicit solvers for the heat equation face a CFL limit on the largest stable time step, $\Delta t < 2/\lambda_N \approx \Delta x^2/2$ for the explicit Euler method, due to the coupling of the temporal and spatial discretizations, so the step count grows as $O(N^2)$ regardless of the accuracy requested.
As such, it is common to employ implicit time integrators to achieve unconditional stability with respect to the time step.
However, implicit time integrators require a linear solve at every step, which creates a computational burden of $O(N^3)$ per step if dense linear solvers are used.
By contrast, \texttt{diff-soup} uses the adaptive VSVO time integrators described in Section~\ref{subsec:vsvo} with the sparse numerical methods of Section~\ref{subsec:sparsity}, which scale quasilinearly in $N$.

\begin{figure}[H]
\centering
\includegraphics[width=\linewidth]{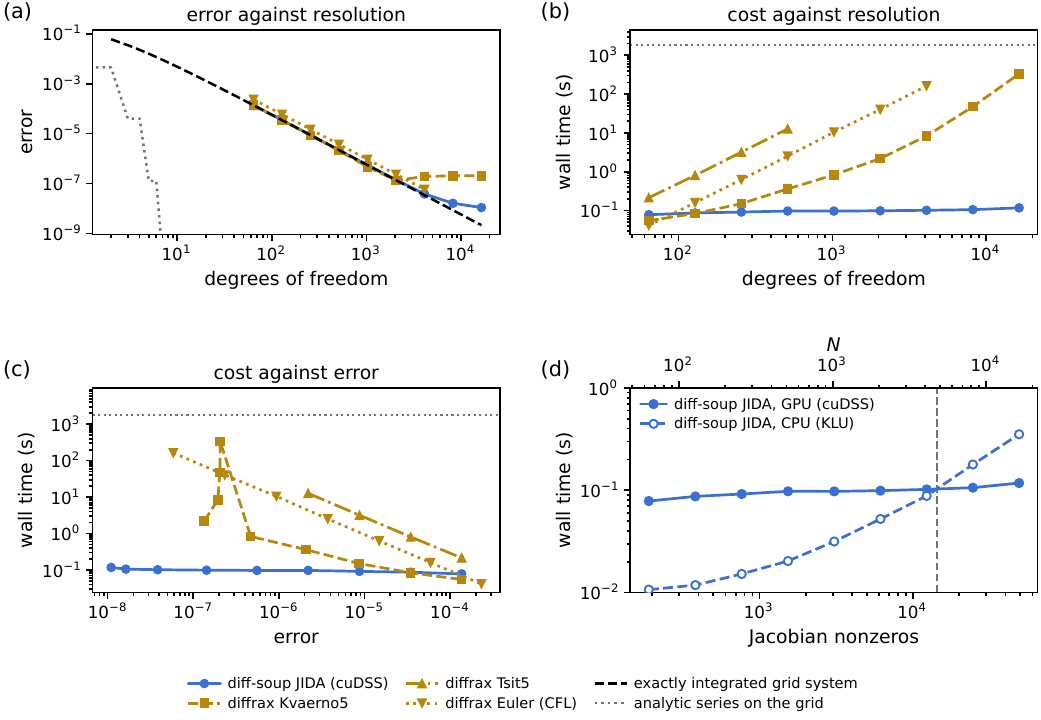}
\caption{
The one-dimensional heat equation solved at $\mathrm{rtol} = \mathrm{atol} = 10^{-9}$ with \texttt{diff-soup} and \texttt{diffrax} on grids from $N = 2^6$ to $2^{14}$ interior nodes.
A maximum wall time of 1800 s is imposed on every timed cell of the benchmark comparisons, indicated by the dotted gray horizontal line.
All problems are solved on the GPU in Float64 precision unless indicated otherwise.
(a) Relative $L^2$ trajectory error against the continuous solution, with the exactly integrated grid system (black dashed) as the $O(\Delta x^2)$ spatial floor and the analytic series truncated to the grid's modes (gray dotted).
At the same tolerance, \texttt{diff-soup} keeps descending to the finest grid and finishes an order of magnitude below \texttt{Kvaerno5}, whose error saturates at the largest sizes.
(b) wall time against resolution.
\texttt{diff-soup} stays flat at a sub-second cost, while the \texttt{diffrax} lanes grow as power laws, the implicit \texttt{Kvaerno5} approaching cubic scaling at the largest sizes.
(c) The work-precision diagram, in which \texttt{diff-soup} gains accuracy at nearly constant cost while every \texttt{diffrax} lane pays a growing cost for each gain in accuracy.
(d) The same \texttt{diff-soup} solver on the CPU with the KLU sparse linear solver and on the GPU with \texttt{cuDSS}, plotted against the Jacobian nonzeros and the degrees of freedom $N$.
The crossover where the CPU cost exceeds the fixed overhead of the GPU solver occurs near $1.4 \times 10^{4}$ nonzeros and 4770 degrees of freedom.
}
\label{fig:heat-1d}
\end{figure}

In Figure~\ref{fig:heat-1d}, we demonstrate that even for this simple parabolic problem, the discrete spectrum of the differential operator induces stiffness that makes the specialized VSVO stiff solver extremely advantageous.
Specifically we find the JIDA integrator with sparse linear algebra more cost-effective than the GPU time integrators \texttt{diffrax} (Figure~\ref{fig:heat-1d}c).
Every method tracks the $O(\Delta x^2)$ spatial floor at the resolutions it reaches, although \texttt{Kvaerno5} saturates near $2 \times 10^{-7}$ from $N = 2^{12}$ onward (Figure~\ref{fig:heat-1d}a).
However, only the \texttt{JIDA} solver of \texttt{diff-soup} holds this accuracy at a low, fixed cost across all resolutions (Figure~\ref{fig:heat-1d}b).

Across this range of resolution, \texttt{diff-soup} exhibits a nearly fixed cost of about $0.1$ s, indicating that the wall time is dominated by the overhead of launching GPU kernels and of host-device synchronization at the branch points of the adaptive integration loop.
To discriminate the performance gains of the GPU accelerator versus the sparse VSVO algorithm, we run the same algorithm on the CPU with the KLU sparse linear solver (Figure~\ref{fig:heat-1d}d).
We note that once the cost is dominated by the sparse linear solver, the precise scaling follows the total cost of the factorization kernels, which depends nontrivially on both the number of nonzeros and the exact sparsity structure of the problem, so we plot the cost against the Jacobian nonzeros rather than the degrees of freedom.
We find the cost of the CPU solver exceeds the GPU solver above approximately $1.4 \times 10^{4}$ nonzero Jacobian entries for this problem.

\subsection{Comparing performance with the Brusselator and Cahn-Hilliard problems}
\label{subsec:bench-nonlinear}
\begin{figure}[H]
\centering
\includegraphics[width=\linewidth]{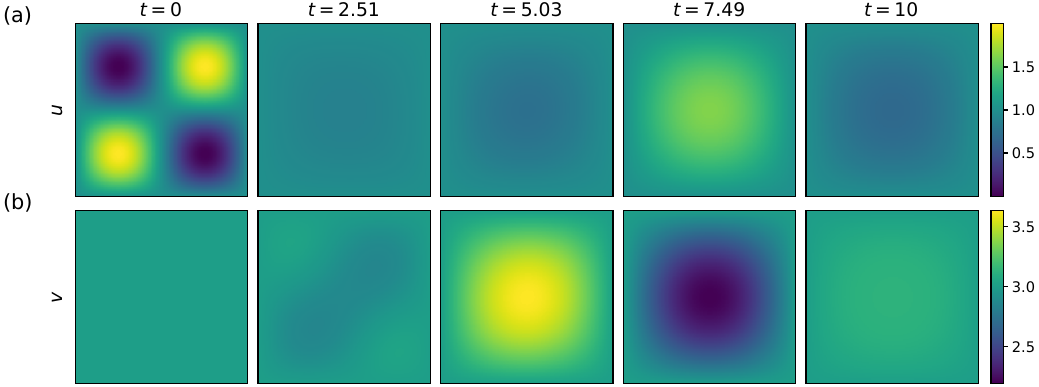}
\caption{
Solution of the two-dimensional Brusselator benchmark, the species $u$ (a) and $v$ (b) on a $128 \times 128$ grid at five times from $t = 0$ to $10$, each row on its own color scale.
}
\label{fig:solution-evolution-brusselator}
\end{figure}

To establish the performance of \texttt{diff-soup}, we benchmark our solvers against comparable scientific computing libraries that offer stiff time integrators, sparse linear solvers, and sensitivity analysis, namely SUNDIALS, PETSc, \texttt{DifferentialEquations.jl}, and \texttt{ode15s} in MATLAB.
We select the two-dimensional Brusselator and a coupled chemo-mechanical Cahn-Hilliard system as model stiff problems.
The Brusselator is a reaction-diffusion model \citep{prigogine_symmetry_1968} that is a conventional benchmark for stiff integrators \citep{hairer_solving_1996,rackauckas_confederated_2019},
\begin{equation}
\begin{aligned}
    \pd{u}{t} &= A + u^2 v - (B+1)\,u + \alpha \nabla^2 u, \\
    \pd{v}{t} &= B\,u - u^2 v + \alpha \nabla^2 v,
\end{aligned}
\label{eq:brusselator}
\end{equation}
where $u$ and $v$ are the concentrations of the two species, $A = 1$ and $B = 3$ are the feed and conversion rates, and $\alpha = 1/50$ is the diffusivity.
The problem is solved on the unit square for $t \in [0, 10]$, with every wall held at the reaction steady state $u = 1$, $v = 3$, from the initial condition $u = 1 + \sin(2\pi x)\sin(2\pi y)$, $v = 3$.
Under these conditions, the system exhibits stiffness from the diffusion term, the nonlinearity, and the oscillatory behavior visible in Figure~\ref{fig:solution-evolution-brusselator}.

\begin{figure}[H]
\centering
\includegraphics[width=\linewidth]{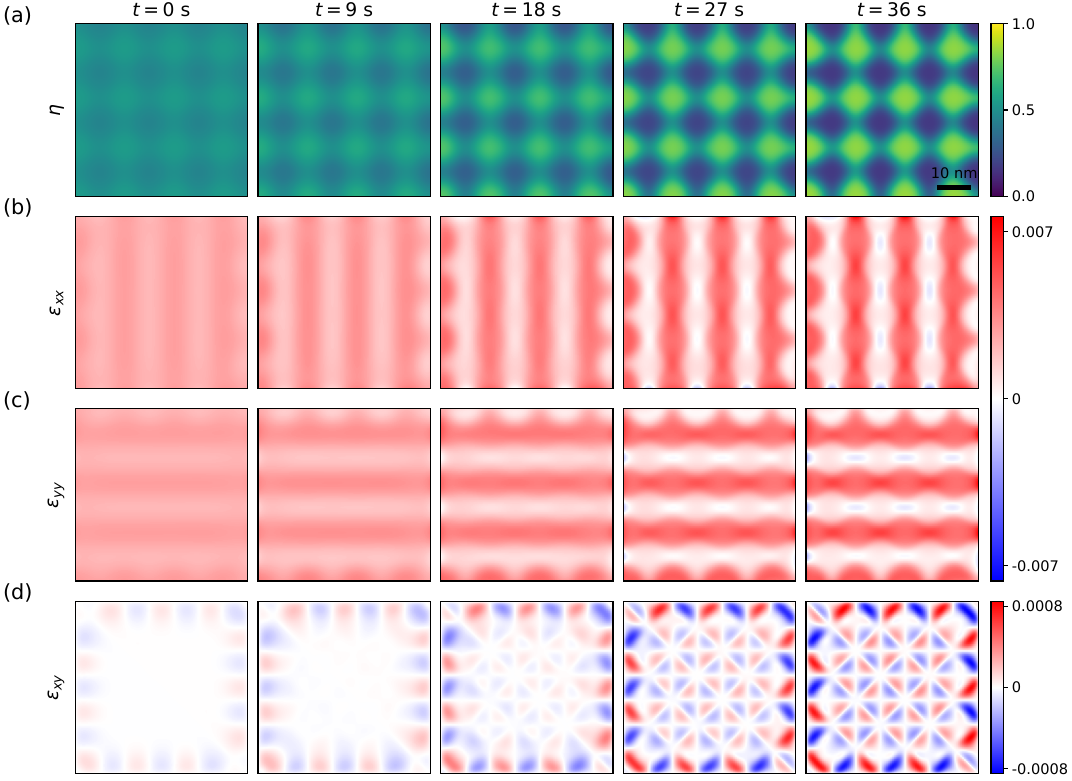}
\caption{
Solution of the Cahn-Hilliard problem with elasticity on its 50 nm square domain at $256 \times 256$ cells, five snapshots from $t = 0$ to $36$ s.
(a) The conserved order parameter $\eta$ phase separating from the sinusoidal seed. 
(b, c, d) The strain components $\varepsilon_{xx}$, $\varepsilon_{yy}$, and $\varepsilon_{xy}$ of the displacement that holds the solid in mechanical equilibrium, the normal components on one shared scale and the shear on its own.
}
\label{fig:solution-evolution-ch}
\end{figure}

To stress-test the solvers, we select the fourth-order Cahn-Hilliard problem coupled to mechanical equilibrium, inspired by the elastic precipitate benchmark of the PFHub phase-field community \citep{jokisaari_phase_2018,wheeler_pfhub_2019}, defined as follows:
\begin{equation}
\begin{aligned}
    \pd{\eta}{t} &= \nabla \cdot \left( M \nabla \mu \right), \\
    0 &= \mu - f_{\mathrm{bulk}}'(\eta) - \pd{f_{\mathrm{el}}}{\eta} + \kappa \nabla^2 \eta, \\
    0 &= \nabla \cdot \sigma, \qquad \sigma = C(\eta) : \left( \varepsilon - \varepsilon^{0}(\eta) \right),
\end{aligned}
\label{eq:ch-elastic}
\end{equation}
where $\eta$ is the conserved order parameter, $\mu$ is its chemical potential, $M$ is the mobility, $f_{\mathrm{bulk}}$ is the bulk free energy, $\kappa$ is the gradient energy coefficient, $\varepsilon = \operatorname{sym}(\nabla \mathbf{u})$ is the strain of the displacement $\mathbf{u}$, $\varepsilon^{0}(\eta)$ is the dilatational misfit strain of the transformed phase, $C(\eta)$ is the stiffness interpolated between the two phases, and $f_{\mathrm{el}}$ is the elastic energy density.
We use an initial condition of two cosine modes with wavelengths near the most unstable wavelength, and integrate the system for $t \in [0, 36]$ s on a $50 \times 50$ nm square domain.
This problem is an index-1 DAE with both local differential and algebraic states, and it exhibits significant stiffness due to the fourth-order spatial derivative, the nonlinearity, and the unstable spatial modes shown in Figure~\ref{fig:solution-evolution-ch}.
The full problem statement and its parameters are given in Section~\ref{subsec:benchmarks}.

\begin{figure}[H]
\centering
\includegraphics[width=\linewidth]{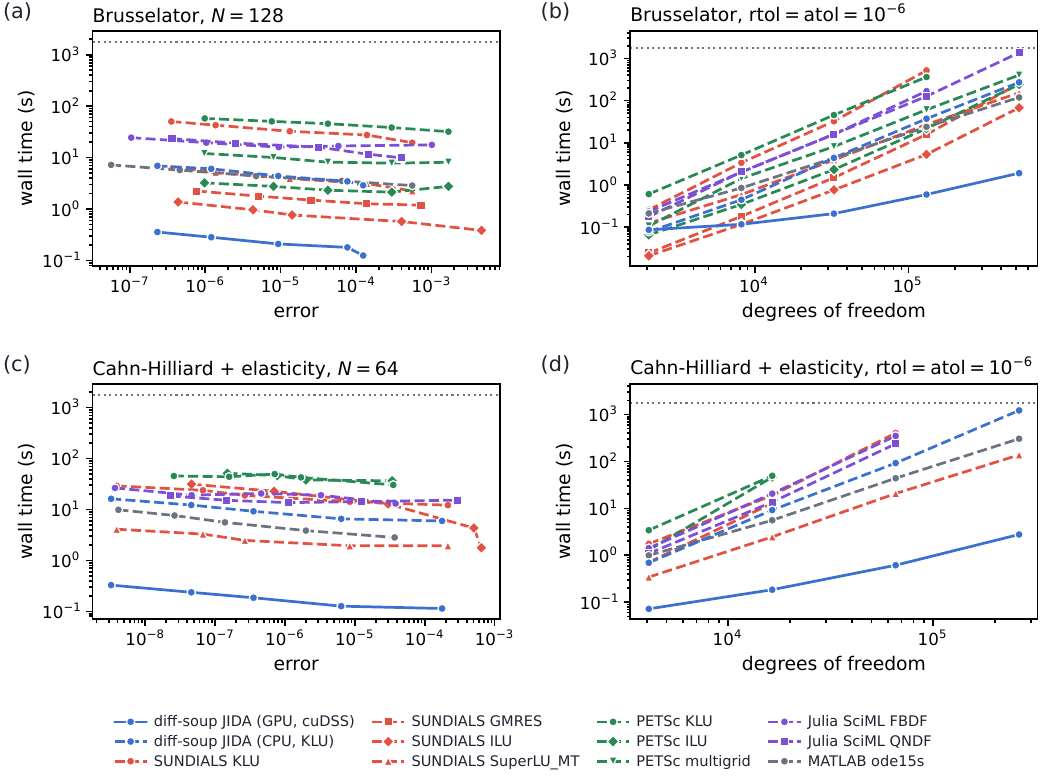}
\caption{
Work-precision and scaling diagrams for the Brusselator (a, b) and the Cahn-Hilliard problem (c, d) for \texttt{diff-soup} on GPU (solid blue) and CPU (dashed blue), SUNDIALS (dashed red), PETSc (dashed green), Julia \texttt{DifferentialEquations.jl} (dashed purple), and MATLAB (dashed gray), with the marker denoting the particular solver as described in the legend.
All benchmarks are run according to the procedure described in Section~\ref{subsec:benchmarks}.
Notably, data points are omitted where the solve did not converge or exceeded the wall cap.
(a) The work-precision diagram for the Brusselator problem shows the \texttt{diff-soup} GPU solver outperforming all solvers, with the Newton-Krylov solvers of SUNDIALS trailing by a small factor and the other libraries by an order of magnitude or more.
(b) The Brusselator scaling diagram shows the \texttt{diff-soup} GPU solver exhibiting a transition from fixed overhead to power-law scaling, whereas the CPU based solvers all exhibit a sharper power-law growth.
(c) The Cahn-Hilliard work-precision diagram shows a similar order of magnitude improvement of the \texttt{diff-soup} GPU solver over the other libraries, with the closest solvers being the sparse direct \texttt{SuperLU\_MT} in SUNDIALS and \texttt{ode15s} in MATLAB.
(d) The Cahn-Hilliard scaling diagram shows the \texttt{diff-soup} GPU solver exhibiting a two order of magnitude improvement over the other libraries at scale, with the CPU based solvers again growing more steeply.
}
\label{fig:forward-cpu-gpu}
\end{figure}

Figure~\ref{fig:forward-cpu-gpu} compares the \texttt{JIDA} solver of \texttt{diff-soup} on the CPU and GPU with the CPU lanes of SUNDIALS, PETSc, \texttt{DifferentialEquations.jl}, and MATLAB on both problems.

As demonstrated in Figure~\ref{fig:forward-cpu-gpu}, the selected benchmarks show the \texttt{diff-soup} GPU solver far outperforms the other solvers, especially above $10^4$ degrees of freedom.
We demonstrate up to a $50\times$ speedup over the next fastest solver at scale.
This performance is explained by using GPUs to overcome the predominant bottlenecks in stiff implicit integration, namely Jacobian computation and factorization of the pencil matrix $W$.

In \texttt{diff-soup}, we employ a parallelized, sparse colored Jacobian computation \citep{curtis_estimation_1974,gebremedhin_what_2005}.
By contrast, all other compared solvers, except \texttt{DifferentialEquations.jl}, must be supplied an analytical sparse Jacobian, as done in this benchmark (Section \ref{subsec:benchmarks}).
\texttt{DifferentialEquations.jl} supports sparse automatic differentiation by Jacobian coloring; however, it evaluates the colors serially and remains slower than the analytical Jacobian computation, as shown in Table~\ref{tab:jacobian-cost}.

\begin{table}[H]
\caption{
Wall time of one residual evaluation, one analytical Jacobian assembly, and one sparse autodiff Jacobian assembly at the work-precision sizes of the two benchmark problems.
Ratios are taken against the analytical Jacobian of the same framework on the same device, and the coloring uses 12 and 67 colors in \texttt{DifferentialEquations.jl} and 14 and 73 in \texttt{diff-soup} for the Brusselator and the Cahn-Hilliard problem, respectively.
}
\label{tab:jacobian-cost}
\centering
\footnotesize
\setlength{\tabcolsep}{5pt}
\begin{tabularx}{\linewidth}{Xlrrrr}
\toprule
& & \multicolumn{2}{c}{Brusselator, $N = 128$} & \multicolumn{2}{c}{Cahn-Hilliard, $N = 64$} \\
\cmidrule(lr){3-4}\cmidrule(lr){5-6}
Computation & Device & Time (ms) & Ratio & Time (ms) & Ratio \\
\midrule
\multicolumn{6}{l}{\emph{Residual evaluation}} \\
\texttt{DifferentialEquations.jl} & CPU & 0.31 & -- & 0.15 & -- \\
\texttt{diff-soup} & CPU & 0.125 & -- & 0.27 & -- \\
\texttt{diff-soup} & GPU & 0.072 & -- & 0.125 & -- \\
\addlinespace
\multicolumn{6}{l}{\emph{Analytical Jacobian}} \\
\texttt{DifferentialEquations.jl} & CPU & 0.46 & -- & 7.3 & -- \\
\texttt{diff-soup} & CPU & 0.37 & -- & 5.0 & -- \\
\texttt{diff-soup} & GPU & 0.052 & -- & 0.154 & -- \\
\addlinespace
\multicolumn{6}{l}{\emph{Sparse autodiff Jacobian}} \\
\texttt{DifferentialEquations.jl}, serial & CPU & 4.6 & 10 & 10.4 & 1.4 \\
\texttt{diff-soup}, serial & CPU & 2.1 & 5.9 & 29 & 5.8 \\
\texttt{diff-soup}, batched & CPU & 0.99 & 2.7 & 30 & 6.0 \\
\texttt{diff-soup}, batched & GPU & 0.12 & 2.4 & 0.35 & 2.3 \\
\bottomrule
\end{tabularx}
\end{table}

On the CPU, the residual and automatic differentiation operations quickly become memory bound, so every color of the serial sweep costs a full residual evaluation, and the sparse autodiff Jacobian runs up to an order of magnitude slower than the analytical one.
Batching offers at most a modest gain, since the cores are already saturated.
To compensate, CPU codes can rely on BDF and NDF algorithms, which generally require fewer Jacobian refreshes than the IDA class of algorithms.
On the GPU, the same batch of colors is evaluated in parallel across the device, which brings the batched \texttt{diff-soup} Jacobian to within a factor of about 2.4 of the analytical Jacobian (Table  \ref{tab:jacobian-cost}), so the factorization of the pencil matrix $W$ remains the bottleneck.

For the two-dimensional problems considered here, the fill-in of the sparse factorization grows superlinearly with the degrees of freedom \citep{george_nested_1973,davis_direct_2006}, so the factorization dominates the solver cost at scale.
Both \texttt{diff-soup} and the alternative scientific computing libraries employ sparse direct LU solvers, however at the scale of large two- and three-dimensional PDEs the superlinear scaling of the factorization becomes prohibitively expensive for CPU based algorithms like \texttt{SuperLU\_MT}, KLU, and the sparse LU of MATLAB, which are limited to a single core or a small number of threads.
\texttt{diff-soup} is able to overcome this on much larger problems because \texttt{cuDSS} distributes the factorization across the GPU and exploits its massive parallelism, achieving up to a $440\times$ speedup compared to the same JIDA integration algorithm using a CPU-based KLU solver.

\begin{figure}[H]
\centering
\includegraphics[width=\linewidth]{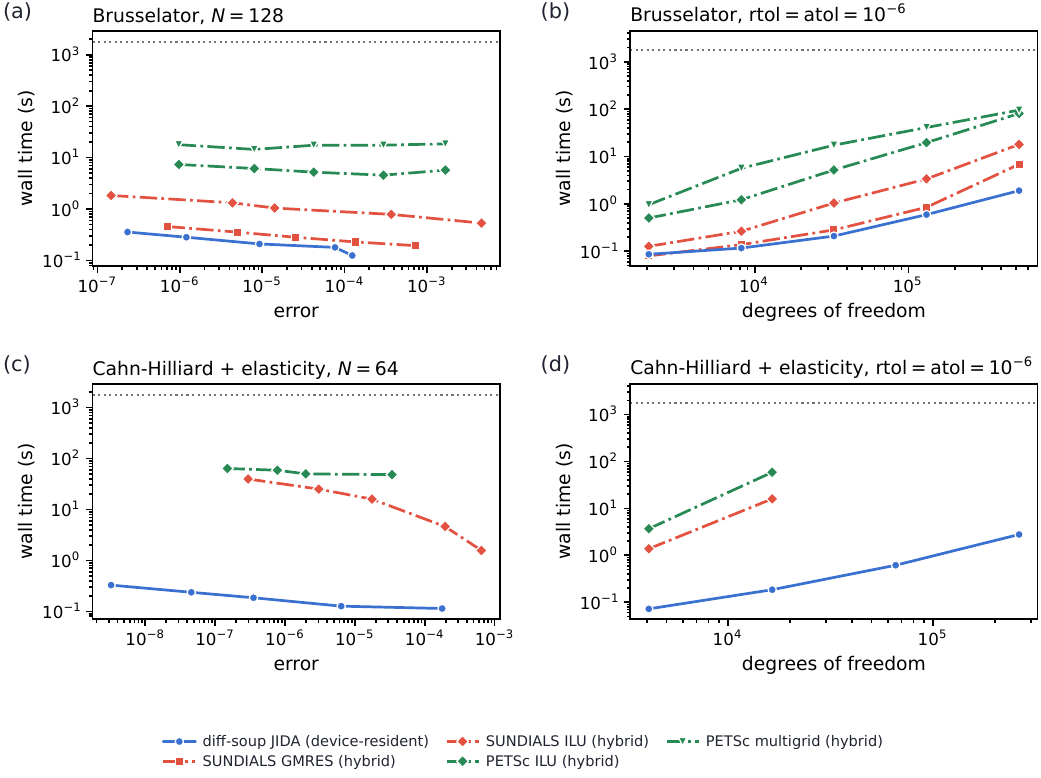}
\caption{
Work-precision and scaling diagrams for the Brusselator (a, b) and the Cahn-Hilliard problem (c, d) for device-resident \texttt{diff-soup} (solid blue) and the hybrid GPU execution of SUNDIALS (dash-dot red) and PETSc (dash-dot green), with the marker denoting the particular solver as described in the legend.
All benchmarks are run according to the procedure described in Section~\ref{subsec:benchmarks}.
Notably, data points are omitted where the solve did not converge or exceeded the wall cap.
(a) The work-precision diagram for the Brusselator problem shows \texttt{diff-soup} outperforming both hybrid solvers, with the unpreconditioned Newton-Krylov solver of SUNDIALS following within a small factor and the multigrid-preconditioned PETSc solver trailing by over an order of magnitude.
(b) The Brusselator scaling diagram shows \texttt{diff-soup} extending its lead over the hybrid SUNDIALS solver as the problem grows, with the PETSc solver trailing by more than an order of magnitude at scale.
(c) The Cahn-Hilliard work-precision diagram shows \texttt{diff-soup} outperforming both hybrid solvers by one to three orders of magnitude, with the closest being the preconditioned Newton-Krylov solver of SUNDIALS.
(d) The Cahn-Hilliard scaling diagram shows only \texttt{diff-soup} reaching the largest sizes, with both hybrid solvers growing far more steeply before they stop.
}
\label{fig:gpu-hybrid}
\end{figure}

However, the GPU based \texttt{diff-soup} code incurs a fixed overhead, since the device kernels sacrifice per-operation efficiency for parallelism and every step requires GPU kernel launches and host-device synchronization.
Consequently, the GPU solver exhibits a nearly constant cost on the Brusselator and Cahn-Hilliard problems until the factorization grows large enough to dominate the wall time (Figure~\ref{fig:forward-cpu-gpu}b and d).
The one-dimensional Brusselator, at sizes too small for the GPU to pay off, is reported in~\ref{app:brusselator-1d}.
To better separate the gains due to the GPU from those due to the sparse direct algorithm, in Figure~\ref{fig:gpu-hybrid} we compare \texttt{diff-soup} against the hybrid GPU execution offered by SUNDIALS and PETSc, in which the time stepper and the Newton iteration remain on the host while the Newton-Krylov kernels execute on the GPU.

Traditionally, Newton-Krylov methods are used to scale CPU based codes to large problems, since they require only Jacobian-vector products rather than materializing and factorizing the full Jacobian \citep{knoll_jacobian-free_2004}.
However, their efficacy is strongly problem dependent, as their performance hinges on a well-tuned preconditioner.
Comparing Figure~\ref{fig:forward-cpu-gpu}a and c, the Newton-Krylov solvers work quite well for the Brusselator, with the standard ILU preconditioner or none at all, since the problem is only mildly nonlinear and stiff.
On the Cahn-Hilliard problem, whose nonlinearity and stiffness are far stronger, the generic preconditioners are much less effective, and the more robust sparse direct solvers perform best.
This trend carries over to our hybrid GPU benchmark in Figure~\ref{fig:gpu-hybrid}, where we find that on the easier problem the Newton-Krylov based SUNDIALS solver closely tracks the performance of \texttt{diff-soup} in both the work-precision and the scaling diagram (Figure~\ref{fig:gpu-hybrid}a and b), but for the harder Cahn-Hilliard problem the Newton-Krylov based solvers fall behind \texttt{diff-soup} by orders of magnitude (Figure~\ref{fig:gpu-hybrid}c and d).
The GPU therefore accelerates a Newton-Krylov method only where the Krylov iteration already converges well, whereas the sparse direct factorization benefits from the device on both problems.
Consequently, we have focused on developing \texttt{diff-soup} around sparse direct methods, as they are generally more robust across a wide range of nonlinear, stiff PDE problems.
However, we note that Newton-Krylov methods are still extremely effective when using problem-specific preconditioners, especially for three-dimensional PDEs where sparse direct methods become memory bound due to the superlinear fill in (Section \ref{sec:GPUaccel}) \cite{george_nested_1973}.
As such, we plan to expand \texttt{diff-soup} in the near future to support preconditioned Newton-Krylov-methods.

\begin{figure}[H]
\centering
\includegraphics[width=\linewidth]{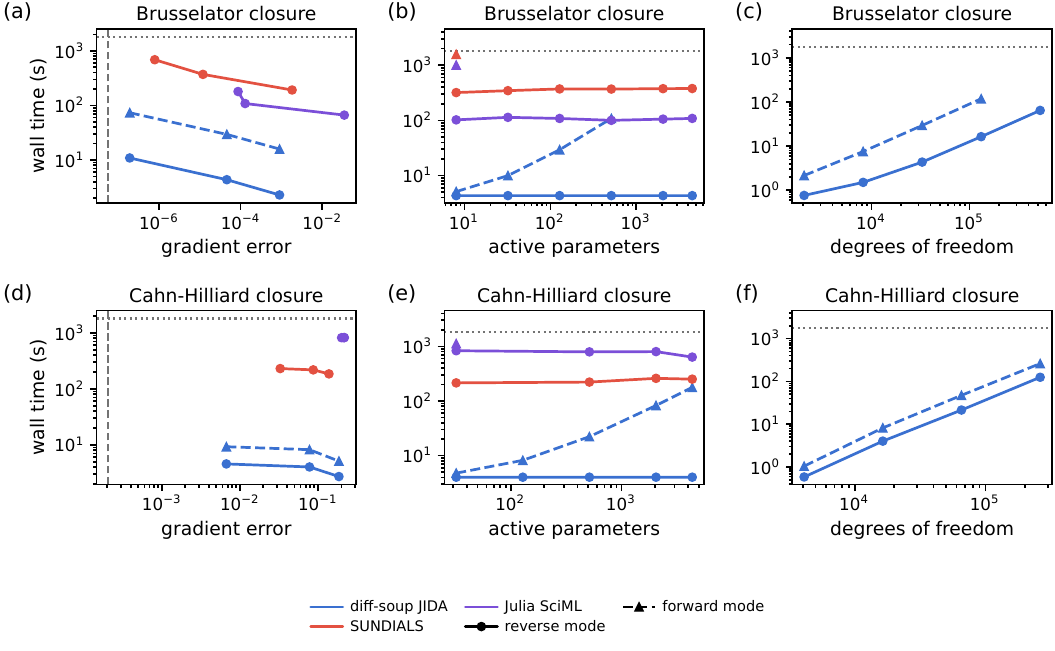}
\caption{
Work-precision and scaling diagrams for the gradient of a loss function with respect to the closure weights of the Brusselator (a, b, c) and the Cahn-Hilliard problem (d, e, f) for \texttt{diff-soup} (blue), the CVODES and IDAS sensitivity methods of SUNDIALS (red), and \texttt{SciMLSensitivity} with \texttt{FBDF} in Julia (purple), with the line style and marker denoting forward or reverse mode as described in the legend.
All benchmarks are run according to the procedure described in Section~\ref{subsec:benchmarks}.
Notably, data points are omitted where the solve did not finish or exceeded the wall cap.
The work-precision diagrams hold the state and the network fixed, the parameter diagrams grow the active weights at fixed tolerance, and the state diagrams grow the degrees of freedom at fixed network width for the \texttt{diff-soup} lanes alone.
The vertical dashed line marks the noise floor of the reference gradient.
(a) The Brusselator work-precision diagram at $N_{\mathrm{dof}} = 32{,}768$ and $n_\theta = 128$ shows \texttt{diff-soup} in reverse mode outperforming the adjoint solvers by one to two orders of magnitude at every tolerance.
(b) The Brusselator parameter scaling diagram shows the reverse-mode cost of every library independent of the number of weights while the forward-mode cost grows with it, with \texttt{diff-soup} alone completing forward mode beyond the smallest network.
(c) The Brusselator state scaling diagram shows both modes of \texttt{diff-soup} growing as a power law in the degrees of freedom, with reverse mode reaching the largest size.
(d) The Cahn-Hilliard work-precision diagram at $N_{\mathrm{dof}} = 16{,}384$ and $n_\theta = 128$ shows a similar order of magnitude improvement, with every library reaching comparable gradient errors.
(e) The Cahn-Hilliard parameter scaling diagram shows the same separation as (b), with forward mode on the CPU completing at most the smallest network.
(f) The Cahn-Hilliard state scaling diagram shows both \texttt{diff-soup} modes growing as a slightly steeper power law and reaching the largest size.
}
\label{fig:gradients}
\end{figure}

Lastly, we compare the performance of differentiating through the forward PDE solver by taking the gradient of a loss function defined on the model output with respect to the model parameters in both forward and reverse mode.
We replace the Brusselator reaction term and the Cahn-Hilliard enthalpic term with pointwise neural networks, which leave the sparsity pattern of the Jacobian unchanged \citep{rackauckas_universal_2021}, to measure the scaling from small classical to large parameter counts.
Here, among the libraries of the forward comparison, only SUNDIALS, through CVODES and IDAS, and \texttt{DifferentialEquations.jl}, through \texttt{SciMLSensitivity}, combine stiff integrators and sparse direct solvers with native sensitivity analysis, since the adjoint of PETSc does not support its BDF time stepper and MATLAB provides no sensitivity analysis for \texttt{ode15s}.
We analyze the cost and accuracy tradeoff, the scaling with respect to the parameter count, and the scaling with respect to the degrees of freedom, as shown in Figure~\ref{fig:gradients}.

As discussed in Section~\ref{subsec:backpdeopt}, the cost of a reverse-mode gradient is one backward sweep independent of the parameter count, $O(T(n) + n_\theta)$, whereas forward mode carries one tangent system per parameter and scales as $O(n_\theta\, T(n))$, and Figure~\ref{fig:gradients}b and e exhibit this separation, with the reverse-mode cost of every library nearly independent of the number of weights and the forward-mode cost growing with it.
Consistent with the forward benchmarks, we find that \texttt{diff-soup} is one to two orders of magnitude faster than the CPU codes at scale, since each step of the backward sweep requires one sparse direct solve with the transposed pencil matrix, which is performed by the same GPU-accelerated factorization as the forward solve.
We measure accuracy in Figure~\ref{fig:gradients}a and d as the relative $L^{2}$ error against the reference gradient of Section~\ref{subsec:benchmarks}, which we validated against central finite differences.
On the Cahn-Hilliard problem, the gradient error of every library remains above $10^{-3}$ at all tolerances, since the unstable spatial modes of Eq.~\eqref{eq:ch-elastic} amplify the local error of the solve into the gradient.
The GPU acceleration also extends the range of network sizes over which forward mode remains practical, owing to the tangent batching of Section~\ref{subsec:backpdeopt}, which assembles all tangents into a single batched multiple right-hand-side solve against the primal factorization, so that \texttt{diff-soup} differentiates networks of hundreds to thousands of weights in forward mode, whereas the CPU codes complete only the smallest network close to the wall cap.
Finally, the checkpointed adjoint bounds the memory of the backward sweep, so reverse mode reaches the largest states of both ladders in Figure~\ref{fig:gradients}c and f, whereas forward mode stores one tangent of the full state per parameter and, with 128 parameters, exhausts the device memory beyond about $10^5$ degrees of freedom on the Brusselator ladder.
As we explore in Section~\ref{subsec:app-activity-field}, these performance advantages enable the training of embedded neural networks at scale within a general-purpose stiff PDE solver.

\subsection{Reduced Precision and Multi-GPU Parallelism}
\label{sec:GPUaccel}
\begin{figure}[H]
\centering
\includegraphics[width=\linewidth]{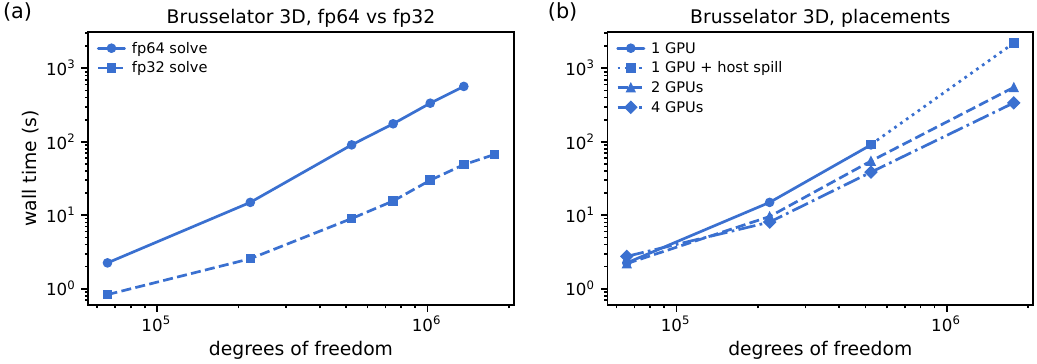}
\caption{
Scaling diagrams for the three-dimensional Brusselator for \texttt{diff-soup} \texttt{JIDA} at $\mathrm{rtol} = \mathrm{atol} = 10^{-6}$ on L40S devices, with the line style denoting the precision or the device placement as described in the legend.
Wall time is measured according to the procedure described in Section~\ref{subsec:benchmarks}, with no cap line because these runs are bounded by the scheduler slice rather than by the wall cap.
Notably, data points are omitted where the factorization did not fit in memory.
(a) The single-device diagram shows Float32 precision outperforming Float64 precision by an order of magnitude at scale and reaching a larger grid.
(b) The placement diagram shows two devices roughly halving the cost of one and four devices improving on two with diminishing returns, with host spill keeping the largest grid solvable on one device at a far higher cost.
}
\label{fig:gpu-workflows}
\end{figure}

Traditionally, sparse direct solvers have struggled to remain competitive with Newton-Krylov methods for large, three-dimensional problems due to the superlinear growth of their computational cost and memory requirements \citep{george_nested_1973,davis_direct_2006}.
However, as discussed in Section~\ref{subsec:bench-nonlinear}, sparse direct methods have significant robustness advantages over Newton-Krylov methods, which has led us to explore GPU-specific strategies for extending sparse direct methods to three-dimensional problems.
Using the three-dimensional Brusselator with the same parameters and wall data as Eq.~\eqref{eq:brusselator}, we test the advantages of reduced precision and multi-GPU model parallelism in Figure~\ref{fig:gpu-workflows}.

For the memory-bound sparse factorization of three-dimensional systems, precision is a significant lever, since many accelerators run Float64 precision at a small fraction of their Float32 precision rate.
As seen in Figure~\ref{fig:gpu-workflows}a, the cost of the Float32 precision solve falls by an order of magnitude at scale, and it prevents the memory overflow of the Float64 precision factor at $1.8 \times 10^{6}$ degrees of freedom.
At matched tolerance, its final state agrees with the Float64 precision solve to within $4 \times 10^{-6}$ in relative $L^{2}$ norm at every smaller size.
Consequently, if the stiffness of the problem tolerates a Float32 precision factor, reduced precision is the most effective acceleration available, especially in three dimensions where the bottleneck is the factorization of the pencil matrix.
In reverse mode, the Float32 precision gradient of the Brusselator closure problem on a $32 \times 32$ grid at $\mathrm{rtol} = 10^{-5}$ agrees with the Float64 precision gradient to a relative $L_2$ error of $10^{-3}$.
Distributing the factorization across devices is a weaker lever, as seen in Figure~\ref{fig:gpu-workflows}b, roughly halving the cost on two devices and improving only modestly on four.
We attribute this to a distributed factorization parallelizing the independent subtrees of the elimination tree while the large separators at the top of the tree are still factored on a single device, which caps the speedup in three dimensions.
The more significant gain is memory capacity, as pooling the devices enables problems that no longer fit on a single GPU, whereas spilling the factor into host memory keeps such problems feasible on one device at a much higher cost.
Even so, the exact factorization eventually exceeds the combined memory of four devices, which sets the practical limit of this approach.

\section{Applications to PDE-constrained Optimization}
\label{sec:apppdeopt}
\texttt{diff-soup} combines physics modeling, differentiable simulation, and optimization in a single package that serves both classical and deep-learning based inference.
In this section, we demonstrate end-to-end workflows implemented entirely in \texttt{diff-soup} for two different inference paradigms: a conventional Bayesian inference problem for learning the constitutive relationships of binary phase-separating materials, inspired by \citet{zhao_image_2021} and \citet{cohen_differentiable_2026}, and a modern deep learning problem in which a neural field is embedded in a packed bed reactor model and trained to represent its three-dimensional catalyst activity.

\subsection{Inference with Uncertainty Quantification}
\label{subsec:app-inverse-ch}

First, we demonstrate \texttt{diff-soup} on a classic Bayesian inverse problem, in which the constitutive relationships of a phase-separating material are inferred from image data.
We reproduce the results of \citet{zhao_image_2021}, inferring the composition-dependent tracer diffusivity $D(c)$ and the homogeneous chemical potential $\mu_h(c)$ from a single noisy movie of spinodal decomposition and quantifying the uncertainty by sampling the posterior.
The composition $c$ evolves according to the Cahn-Hilliard equation
\begin{equation}
\begin{aligned}
    \pd{c}{t} &= \nabla \cdot \big( D(c)\, c\, \nabla \mu \big), \\
    \mu &= \mu_h(c) - \kappa \nabla^2 c,
\end{aligned}
\label{eq:we1-ch}
\end{equation}
where $\mu$ is the chemical potential in units of $k_B T$ per particle, $\kappa$ is the gradient energy coefficient, and $D(c)$ is the tracer diffusivity, related to the mobility $M(c) = c\,D(c)$ by the Einstein relation.
We parameterize both functions with orthogonal polynomials,
\begin{equation}
    \mu_h(c) = \ln \frac{c}{1-c} + \sum_{k=1}^{12} a_k P_k(c),
    \qquad
    D(c) = \exp\Bigl( \sum_{k=0}^{9} b_k P_k(c) \Bigr),
    \label{eq:we1-laws}
\end{equation}
where $P_k$ is the shifted Legendre polynomial of degree $k$ normalized in \(L^{2}([0,1])\) , the logarithmic term retains the ideal entropy of mixing, the exponential keeps the diffusivity positive, and $\theta = (a, b)$ collects the coefficients.
Following \citet{zhao_image_2021}, we assume the coefficients are distributed as
\begin{equation}
    a_k,\, b_k \sim \mathcal{N}\!\left(0,\, \frac{s^2}{1 + k(k+1)}\right),
    \label{eq:we1-prior}
\end{equation}
where the variance decays with the polynomial degree, so that the prior damps oscillations that the data cannot resolve, and the scale $s$ is set by empirical Bayes (Section~\ref{subsec:we1}).
We then solve for the maximum a posteriori (MAP) estimate
\begin{equation}
    \theta^\ast = \operatorname*{arg\,min}_{\theta} \; \frac{1}{2\sigma^2} \sum_{i=1}^{N} \bigl\| c(t_i;\theta) - \hat c(t_i) \bigr\|^2 - \ln p(\theta),
    \label{eq:we1-posterior}
\end{equation}
where $\hat c(t_i)$ is the measured frame at time $t_i$, $\sigma$ is the standard deviation of the measurement noise, and $p(\theta)$ is the prior of Eq.~\eqref{eq:we1-prior}.
We use multiple shooting, so that each prediction $c(t_i;\theta)$ is integrated from the previous measured frame $\hat c(t_{i-1})$ rather than from the start of the movie, which keeps each solve short and greatly accelerates the optimization \citep{bock_multiple_1984}.
As in \citet{cohen_differentiable_2026}, we treat each measured frame as an exact initial state for its window, so the likelihood of Eq.~\eqref{eq:we1-posterior} accounts for the noise in the predicted frame but not in the initial one.
We solve Eq.~\eqref{eq:we1-posterior} with the Newton trust region algorithm and then sample the posterior with the No-U-Turn sampler (NUTS) \citep{hoffman_no-u-turn_2014} implemented in \texttt{blackjax} \citep{cabezas_blackjax_2024}.

\begin{figure}[H]
\centering
\includegraphics[width=\linewidth]{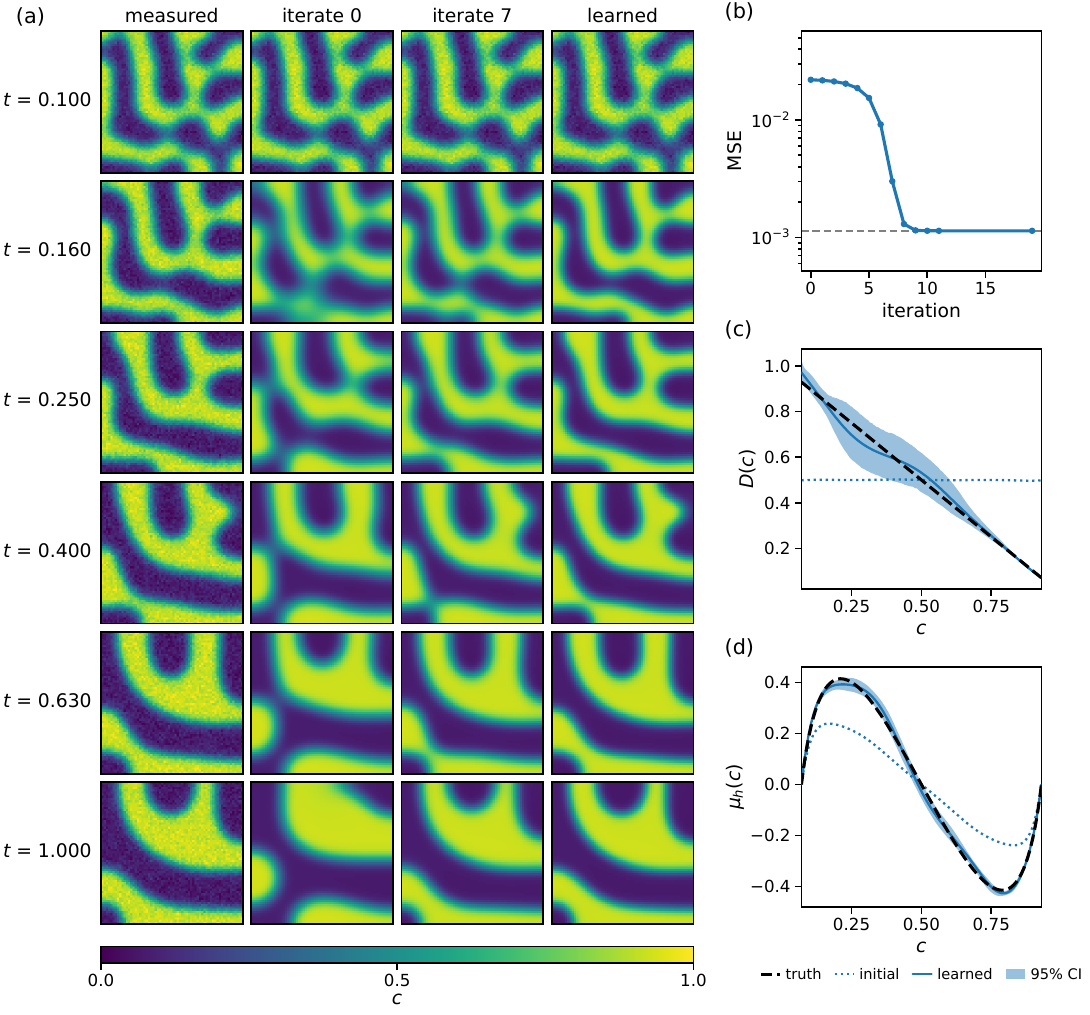}
\caption{
Recovery of the diffusivity $D(c)$ and the homogeneous chemical potential $\mu_h(c)$ of a Cahn-Hilliard model from one noisy movie of spinodal decomposition at the 10\% measurement noise level.
(a) The dynamical trajectory as measured (left) and as predicted by the model at iterate 0, at iterate 7, and at the final iterate.
(b) Mean squared error in $c$ against trust-region iteration, converging to the measurement noise floor (gray dashed).
(c, d) The learned laws (solid) and the initial laws (dotted) against the ground truth (black dashed) with the 95\% credible interval (CI) of the NUTS posterior.
}
\label{fig:we1-inverse-ch}
\end{figure}

Figure~\ref{fig:we1-inverse-ch} summarizes the results.
The learned model reproduces the dynamics of the noisy movie, as shown in panel (a).
The Newton trust region algorithm converges rapidly, reaching the noise level of the measurement within ten iterations in panel (b).
The inferred functions in panels (c) and (d) lie very close to the ground truth, and the posterior samples quantify their uncertainty through 95\% credible intervals.
During spinodal decomposition, the measured compositions lie predominantly near the two equilibrium values and only rarely inside the spinodal gap ($0.21 < c < 0.79$), so the learned functions in this region are informed primarily by the prior and carry a considerably larger uncertainty.

\subsection{Embedded Deep Learning Model}
\label{subsec:app-activity-field}

\texttt{diff-soup} provides the capability to embed large deep learning models directly into PDE simulations, owing to its foundations in \texttt{JAX} and its GPU acceleration.
In this example, we demonstrate physics-constrained deep learning at scale by training a neural network through the solution of a three-dimensional packed bed reactor model, in which the network represents the heterogeneous catalyst activity $k(\mathbf{x})$.
The concentration $c$ of the reactant evolves according to the advection-dispersion-reaction equation
\begin{equation}
    \pd{c}{t} + \pd{}{x}\bigl( u_x(y, z)\, c \bigr) = \nabla \cdot \bigl( \mathbf{D} \nabla c \bigr) - k(\mathbf{x})\, \frac{c}{K + c},
    \label{eq:we3-reactor}
\end{equation}
where the reactant enters the bed at the concentration $c_{\mathrm{in}}$, $u_x(y, z)$ is a fixed velocity profile along the axial coordinate $x$, $\mathbf{D}$ carries the axial and radial dispersion coefficients, and $K$ is the saturation constant of the Langmuir-Hinshelwood rate law.
We generate a synthetic dataset by solving Eq.~\eqref{eq:we3-reactor} on a $96 \times 24 \times 24$ grid with a known activity field and adding Gaussian noise. 
The true activity is a log-normal random field multiplied by a poisoning front and a streak of low-activity pellets, which together form a challenging heterogeneous target for the reconstruction.
We parameterize the activity field $k(\mathbf{x})$ with a wavelet-based neural field (WIRE) \citep{saragadam_wire_2023} consisting of $7.5 \times 10^{4}$ parameters.
The final layer is a softplus function to ensure positivity, and we initialize the network to match the uniform nominal activity $\bar k$.
Since the neural field depends only on position, it enters the residual as a field and leaves the sparsity pattern of the Jacobian unchanged.
We train the neural field by minimizing the mean squared error between the simulated and measured concentrations using the AdamW optimizer \citep{loshchilov_decoupled_2019} with a weight decay of $10^{-4}$ and a peak learning rate of $10^{-3}$ under a linear warmup and cosine decay schedule.

\begin{figure}[H]
\centering
\includegraphics[width=0.72\linewidth]{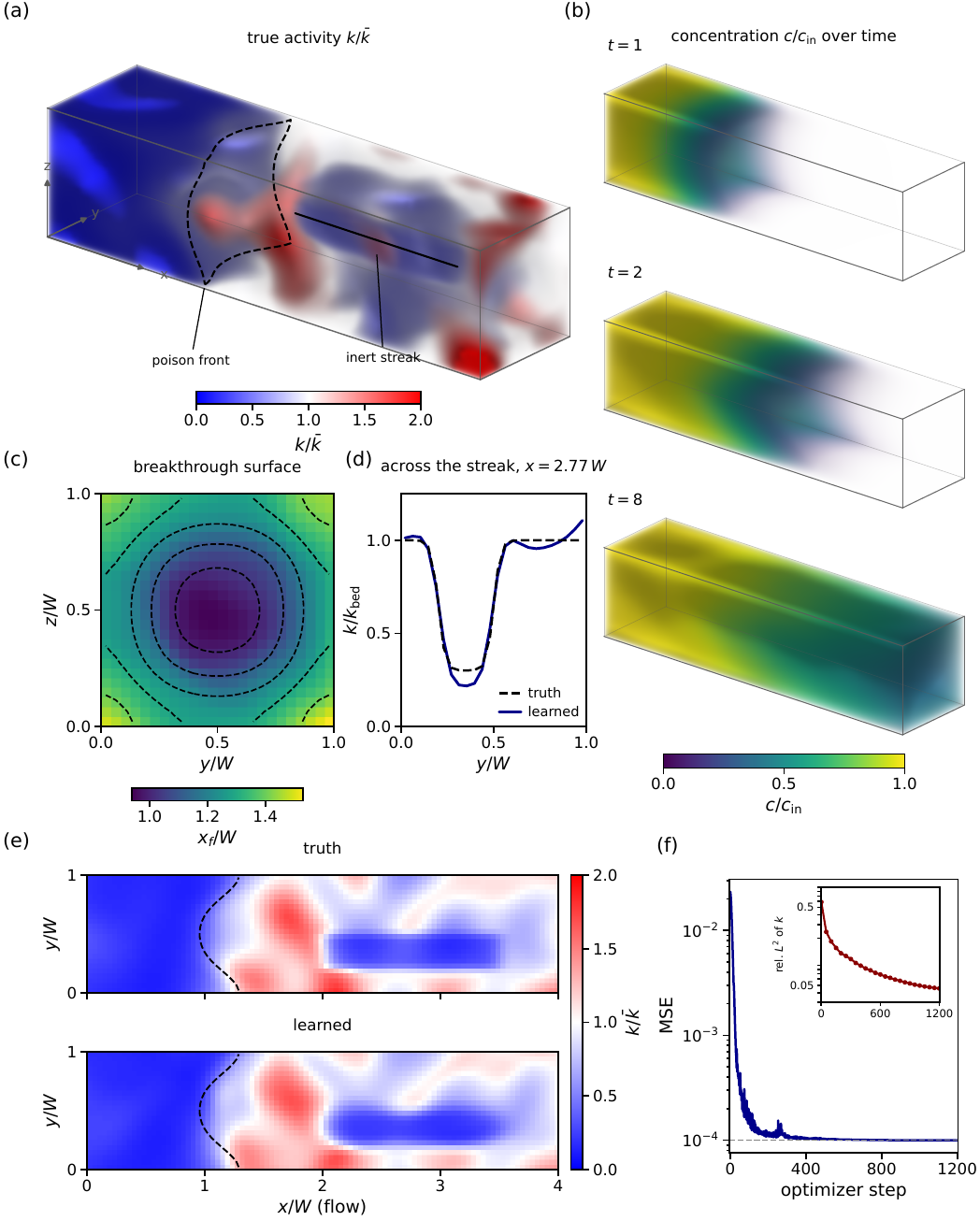}
\caption{
Recovery of a three-dimensional catalyst activity field from noisy concentration measurements in a packed bed reactor.
(a) The true activity $k/\bar{k}$, with the breakthrough surface of the poisoning front (dashed) and the axis of the low-activity streak (solid) marked.
(b) The measured concentration $c/c_{\mathrm{in}}$ at three snapshot times.
(c) The breakthrough surface $x_f(y, z)$ of the poisoning front, as learned (color) and in the ground truth (contours).
(d) The activity across the low-activity streak, normalized by the surrounding bed, as learned (solid) against the ground truth (dashed).
(e) The true and the learned activity on an axial plane through the streak.
(f) Mean squared error in $c$ against optimizer step, converging to the measurement noise floor (gray dashed), with the relative $L^2$ error of the activity in the inset.
}
\label{fig:we3-activity-field}
\end{figure}

Figure~\ref{fig:we3-activity-field} shows the problem and the learned reconstruction.
The training converges rapidly, reaching the noise level of the measurement within a few hundred steps in panel (f).
The learned field recovers the heterogeneity of the true field, capturing the breakthrough surface of the poisoning front in panel (c), the streak of low-activity pellets in panel (d), and the large-scale texture of the random field in panel (e), with a relative $L^2$ error in $k$ of about 5\%.
We note that the remaining error is concentrated in the fine-scale features of the field, since dispersion smooths them out of the measured concentration.
The full training of 1200 steps required a little over four hours on a single NVIDIA L40S GPU.

\section{Conclusion}
\label{sec:conclusion}
We have presented \texttt{diff-soup}, a software package for GPU-accelerated simulation and inference of large-scale dynamical systems built around the cuDSS sparse direct solver.
Accelerating the linear solves of stiff implicit integration on the GPU yields up to $50\times$ faster simulations and $80\times$ faster gradients than established CPU libraries.
Nonlinear PDEs on structured and unstructured domains are expressed in a general-purpose modeling system, integrated as stiff ODEs and DAEs in \texttt{JAX}, and differentiated end to end through the discretize-then-optimize approach.
The resulting gradients enable PDE-constrained optimization for both classical and modern inference and learning problems.
Our methods were evaluated in a systematic benchmark suite against SUNDIALS, PETSc, \texttt{DifferentialEquations.jl}, MATLAB, and \texttt{diffrax} on the heat equation, the Brusselator, and a chemo-mechanical Cahn-Hilliard problem, for both forward solves and gradients.
We demonstrated these workflows by inferring the constitutive relationships of a phase-separating material with uncertainty quantification and by training a neural field embedded in a three-dimensional packed bed reactor model.

This work has opened up several interesting directions for future study.
Specifically, as PDE solvers advance and large-scale physics foundation models emerge, there is a critical need for more rigorous benchmarking of stiff PDEs on both regular and irregular geometries.
On the software side, we plan to improve the high-performance computing (HPC) compatibility and acceleration of \texttt{diff-soup} with more thorough support for large, distributed simulation and optimization problems.
Developing better multi-GPU parallelism strategies for sparse matrix factorization is a promising and important direction of future research.
Furthermore, we note the need for support of our algorithms on general accelerator devices and for a modular automatic differentiation library.
Finally, we plan to generalize the inference module to adjacent applications of PDE-constrained optimization, such as model predictive control and design of experiments, and to constrained optimization algorithms like \texttt{IPOPT} \citep{wachter_implementation_2006}.

By modernizing the PDE modeling, simulation, and inference stack, we aim to build an open-source ecosystem for dynamical systems that advances engineering and drives innovation in AI for science.

\section{Methods}
\label{sec:methods}

\subsection{Benchmarking}
\label{subsec:benchmarks}
The benchmarks of Section~\ref{sec:benchmarks} solve the heat equation, the Brusselator reaction-diffusion system, and the Cahn-Hilliard problem with elasticity.
All problems are posed on Cartesian domains, the heat equation discretized by finite differences as in Section~\ref{subsec:bench-heat} and the other two by the finite-volume method.
The Brusselator adapts the constants of the BRUSS problem of \citet{hairer_solving_1996}.
The Cahn-Hilliard problem follows the elastic precipitate benchmark of PFHub \citep{wheeler_pfhub_2019} with a regular-solution bulk free energy, $f_{\mathrm{bulk}} = A\,[\eta \ln \eta + (1 - \eta) \ln (1 - \eta)] + \omega\, \eta (1 - \eta)$ with $A = 0.16$ and $\omega = 0.40$, a gradient energy coefficient $\kappa = 0.29$ and a mobility $M = 5.0$.
The misfit strain is a dilatational strain of 0.005, and the cubic stiffness constants are $(C_{1111}, C_{1122}, C_{1212}) = (250, 150, 100)$ in the matrix and 1.1 times those in the precipitate, both interpolated between the phases by $h(\eta) = \eta^{3}(6\eta^{2} - 15\eta + 10)$.
The initial condition adds two zero-flux cosine modes of index 7 with peak amplitude 0.05 to a uniform composition of 0.5 on a 50~nm square domain.
The boundaries are no-flux in $\eta$ and $\mu$ and traction free, and three displacement degrees of freedom at two corner cells are pinned to remove the rigid-body modes.
In the gradient benchmark, the replaced constitutive relationship in both systems is a pointwise $\tanh$ network with two hidden layers of width 128, initialized by LeCun normal initialization \citep{lecun_efficient_1998} and fitted with Adam to the term it replaces.
Its $N_\theta$ active weights are drawn from each layer in proportion to its size, the remaining weights are held fixed, and $\mathcal{L}$ is the mean squared misfit at eight equally spaced times to target trajectories generated with the weights perturbed by 1\%.

Every benchmark runs on nodes of one type in an HPC cluster, with Intel Xeon Platinum 8562Y+ processors and NVIDIA L40S GPUs.
In the forward benchmarks, the external comparators run each solve on eight CPU cores, except PETSc, which runs on one, and their hybrid GPU lanes on one L40S and one CPU core.
The \texttt{diff-soup} lanes of Figures~\ref{fig:heat-1d}, \ref{fig:forward-cpu-gpu}, \ref{fig:gpu-hybrid} and \ref{fig:gradients} run \texttt{diff-soup} 1.0.0 with \texttt{cuDSS} 1.0.0.35 and \texttt{JAX} 0.11.2, with the CPU lanes on KLU over eight cores (one core in Figure~\ref{fig:heat-1d}); Figure~\ref{fig:gpu-workflows} and the appendix figures were measured with earlier development versions of the same integrators, and the deposited records provide the benchmark data.
The comparators run SUNDIALS 7.8.0, PETSc 3.25.3, Julia 1.12.6 with \texttt{OrdinaryDiffEq} 7.8.1 and \texttt{SciMLSensitivity} 7.119.13, MATLAB R2025b, and \texttt{diffrax} 0.7.0.
The full manifest in the companion repository provides the configuration of every benchmark.
SUNDIALS \citep{hindmarsh_sundials_2005,gardner_enabling_2022}, PETSc \citep{balay_efficient_1997,abhyankar_petscts_2018} and MATLAB lack a native automatic differentiation library for their integrators, so they are supplied the analytical sparse Jacobian of each problem.
\texttt{DifferentialEquations.jl} \citep{rackauckas_differentialequationsjl_2017} assembles its Jacobian by sparse automatic differentiation with \texttt{ForwardDiff.jl} and graph coloring.
Reported wall time is the median of the warm calls of one process after one cold call, and we employ a strict wall cap of 1800~s for each benchmark cell to minimize the computational burden of the large benchmark suite.
Accuracy is the mean absolute error of the final state against the reference, and for a gradient the relative $L^{2}$ error against the reference gradient.
The Brusselator references are external \texttt{CVODE} solves (SciPy \texttt{Radau} in one dimension) and the Cahn-Hilliard references are \texttt{diff-soup} solves, each at tolerances tightened beyond the plotted range and validated by agreement with a second solve at a neighbouring tolerance.
The reference gradients are \texttt{diff-soup} reverse-mode gradients at tightened tolerances, validated against a second solve and against central finite differences.
Independent SUNDIALS solves reproduce the \texttt{diff-soup} references, with \texttt{IDA} matching the Cahn-Hilliard reference to $6 \times 10^{-13}$ and the \texttt{CVODES} and \texttt{IDAS} adjoints matching both reference gradients well below every plotted error.

SUNDIALS integrates the Brusselator with \texttt{CVODE} and the Cahn-Hilliard problem with \texttt{IDA}, on the KLU \citep{davis_algorithm_2010} and \texttt{SuperLU\_MT} \citep{demmel_asynchronous_1999} direct solvers and on GMRES \citep{saad_gmres_1986}, with and without the ILU(0) preconditioner \citep{saad_iterative_2003} written for this study.
PETSc integrates with \texttt{TSBDF} at orders up to five with a lagged Jacobian and preconditioner, and solves the Newton systems with KLU, with GMRES and ILU(0), and with algebraic multigrid.
For the Newton-Krylov methods, we tune the Krylov dimension above the library defaults, to 50 on the Brusselator and 60 on the Cahn-Hilliard problem in SUNDIALS and in the ILU(0) lanes of PETSc, and to 100 in the multigrid lanes of PETSc.
\texttt{DifferentialEquations.jl} integrates with \texttt{FBDF} and \texttt{QNDF} on KLU and differentiates through \texttt{SciMLSensitivity} with a Gauss adjoint in reverse mode and \texttt{ForwardDiff} in forward mode.
The gradient lanes run Julia on one thread, with 8 BLAS threads for the Cahn-Hilliard problem and 16 for the Brusselator.
MATLAB runs \texttt{ode15s} \citep{shampine_matlab_1997} with the hand-written sparse Jacobian and library defaults otherwise.
\texttt{diffrax} \citep{kidger_neural_2021} runs \texttt{Kvaerno5} and \texttt{Tsit5} under its default controller and explicit Euler at the CFL limit in Figure~\ref{fig:heat-1d}.
The manifests, the per-lane settings and command lines, the reference tolerances, the step budgets and the record of every cell are in the companion repository named in the data availability statement.

\subsection{Worked Example 1}
\label{subsec:we1}
The movie of Figure~\ref{fig:we1-inverse-ch} was generated by solving Eq.~\eqref{eq:we1-ch} with the regular-solution laws $\mu_h(c) = \ln[c/(1 - c)] + 3\,(1 - 2c)$ and $D(c) = 1 - c$ on the unit square at $64 \times 64$ cells, with $\kappa = 0.002$, no flux and no wetting at the walls, and an initial state of uniform composition 0.5 plus seeded noise.
Six frames at geometrically spaced times from $t = 0.1$ to 1.0 were corrupted with Gaussian noise at 10~\% of the standard deviation of the noise-free movie and clipped to $[0.01, 0.99]$ so that the logarithmic term stays finite.
The prior scale $s$ was set by empirical Bayes, scanning the Laplace approximation to the evidence at the fit over six values of $s$ and taking the vertex of the quadratic through the three largest, which gave $s = 1.14$.
Every window was solved with \texttt{JIDA} and \texttt{cuDSS} in single precision with forward-mode sensitivities, at $\mathrm{rtol} = \mathrm{atol} = 10^{-4}$ in the fit and $10^{-6}$ in the sampler, and the fit took 37~s on one NVIDIA A100, compilation included.
The posterior was sampled in four replicates of four chains, each for 200 warmup and 100 sampling transitions, with the chains whitened by the Laplace approximation at the fit, at 1.6 to 1.7~h per replicate.
All sixteen chains mixed, with a split $\hat R$ below 1.01 and no divergent transition, and the credible bands of Figure~\ref{fig:we1-inverse-ch} are the pointwise 95~\% intervals of the two laws over the 1,600 draws.
The data generation, the fit and the sampler scripts are in the companion repository named in the data availability statement.

\subsection{Worked Example 2}
\label{subsec:we3}
\begin{figure}[H]
\centering
\includegraphics[width=0.53\linewidth]{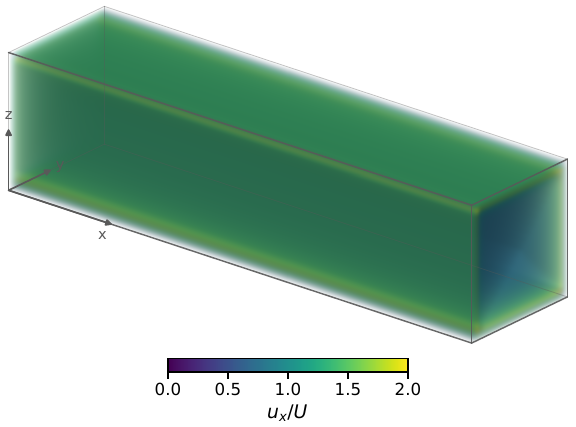}
\caption{
The superficial velocity profile $u_x(y, z)$ of the packed bed from Eq.~\eqref{eq:we3-brinkman}, which channels the flow along the walls.
}
\label{fig:we3-brinkman}
\end{figure}

The reactor of Figure~\ref{fig:we3-activity-field} is a three-dimensional packed bed of dimensions $4W \times W \times W$, non-dimensionalized by the bed width $W$, the mean superficial speed $U$ and the inlet concentration $c_{\mathrm{in}}$.
The velocity profile $u_x(y, z)$ is the steady solution of the Brinkman equation on the cross-section,
\begin{equation}
    \mu_{\mathrm{eff}}\, \nabla^2 u_x - \frac{\mu}{\kappa_p(y, z)}\, u_x = \pd{p}{x},
    \qquad u_x = 0 \text{ on the walls},
    \label{eq:we3-brinkman}
\end{equation}
with the Kozeny-Carman permeability $\kappa_p = d_p^2\, \varepsilon^3 / [150\, (1 - \varepsilon)^2]$ and the Vortmeyer-Schuster wall porosity $\varepsilon(\delta) = \varepsilon_b + (1 - \varepsilon_b)\, e^{-N \delta / d_p}$ at the distance $\delta$ from the nearest wall, for a particle diameter $d_p = W/6$, a bulk porosity $\varepsilon_b = 0.4$, $N = 5$ and $\mu_{\mathrm{eff}}/\mu = 8$, scaled to unit mean (Figure~\ref{fig:we3-brinkman}).
The dispersion coefficients follow from particle P\'eclet numbers of 2 axially and 10 radially, the reactant enters through a Danckwerts flux condition at the inlet and leaves with zero gradient at the outlet, the walls carry no flux, and advection is discretized with central fluxes.
The reaction uses $\bar k = 0.22\,U/W$ and $K = 0.05\,c_{\mathrm{in}}$.

The true activity is a log-normal random field with correlation length $0.25\,W$ and standard deviation 0.3 in $\ln k$, multiplied by a smooth poisoning front that follows the velocity profile and by a cylindrical streak of low-activity pellets at 30\% of the surrounding activity.
The dataset was generated from an empty bed, saved at every cell at eight times from $t = 1$ to 8 and corrupted with Gaussian noise of standard deviation $0.01\,c_{\mathrm{in}}$.
The WIRE network has three hidden layers of width 192 with wavelet parameters $\omega_0 = 10$ and $s_0 = 5$.
The network was trained for 1200 steps on one NVIDIA L40S with the AdamW optimizer \citep{loshchilov_decoupled_2019} in \texttt{optax} \citep{deepmind_deepmind_2020}, every step integrating the model as a single trajectory from $t = 0$ with \texttt{JNDF} and \texttt{cuDSS} in Float32 precision at $\mathrm{rtol} = 10^{-5}$ and $\mathrm{atol} = 10^{-6}$ and using the discrete checkpointed adjoint.
The model, the truth, the data generation and the training scripts are in the companion repository named in the data availability statement.

\section*{CRediT authorship contribution statement}
\textbf{Samuel Degnan-Morgenstern:} Conceptualization, Methodology, Software, Validation, Formal analysis, Investigation, Data curation, Writing - original draft, Writing - review \& editing, Visualization, Project administration, Funding acquisition.
\textbf{Martin Z. Bazant:} Writing - review \& editing, Supervision, Project administration, Funding acquisition.

\section*{Declaration of competing interest}
The authors declare that they have no known competing financial interests or personal relationships that could have appeared to influence the work reported in this paper.

\section*{Acknowledgements}
S.D.M. acknowledges support from the National Science Foundation Graduate Research Fellowship under Grant No. 2141064.
The authors acknowledge the MIT Office of Research Computing and Data for providing high performance computing resources that have contributed to the research results reported within this paper.
The authors also thank Chihyun Nam, Alexander Cohen, Jacob Sass, Sungho Shin, Steven Johnson, and Simon Daubner for valuable discussions.

\section*{Data availability}
The source code of \texttt{diff-soup} is deposited in the CPC Program Library and developed at \url{https://github.com/stmorgenstern/diff-soup}.
The benchmark suite, manifests, result stores, figures, and the scripts of the two worked examples are deposited in the CPC Program Library with the source code.

\section*{Declaration of generative AI and AI-assisted technologies in the writing process}
\label{sec:dec_ai_writing}
During the preparation of this work the authors used Claude Code (Anthropic) in order to assist in software generation and manuscript editing.
After using this service, the authors reviewed and edited the content as needed and take full responsibility for the content of the published article.

\appendix

\section{The three \texttt{diff-soup} VSVO solvers}
\label{app:diffsoup-classes}

\begin{figure}[H]
\centering
\includegraphics[width=\linewidth]{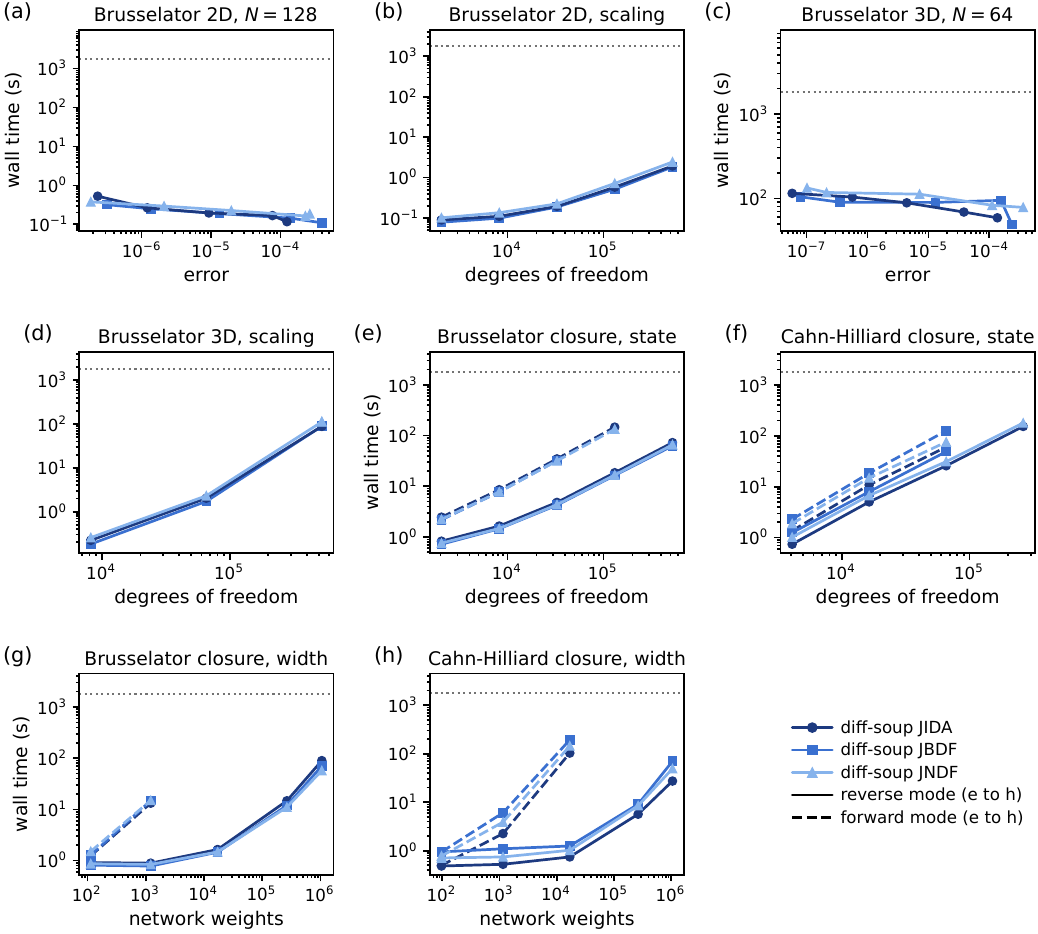}
\caption{
Work-precision and scaling diagrams for the two-dimensional (a, b) and three-dimensional (c, d) Brusselator and gradient scaling diagrams for the Brusselator (e, g) and Cahn-Hilliard (f, h) closures for the \texttt{diff-soup} solvers \texttt{JIDA} (navy), \texttt{JBDF} (blue), and \texttt{JNDF} (light blue), with the marker denoting the solver and the line style the differentiation mode as described in the legend.
All benchmarks are run according to the procedure described in Section~\ref{subsec:benchmarks}.
Notably, data points are omitted where the solve did not converge or exceeded the wall cap.
(a, c) The work-precision diagrams show the three solvers sharing one performance-accuracy frontier.
(b, d) The scaling diagrams show the three solvers exhibiting the same asymptotic scaling within a small factor of one another.
(e, f, g, h) The gradient diagrams show the three solvers within a small factor of one another in both differentiation modes, with the mass-matrix solvers \texttt{JBDF} and \texttt{JNDF} marginally faster on the Brusselator closure and \texttt{JIDA} the fastest on the Cahn-Hilliard closure.
}
\label{fig:diffsoup-classes}
\end{figure}

In this appendix, we compare the three native VSVO solvers of \texttt{diff-soup} on each benchmark problem of Section~\ref{sec:benchmarks}.
Figure~\ref{fig:diffsoup-classes} shows that the three solvers generally exhibit similar performance at matched error, with no significant leader.
For this reason, we adopt \texttt{JIDA} for the main text results, as it is generally the most robust stiff implicit integration algorithm and handles the most general problem form.

\section{One-dimensional Brusselator comparison}
\label{app:brusselator-1d}

\begin{figure}[H]
\centering
\includegraphics[width=\linewidth]{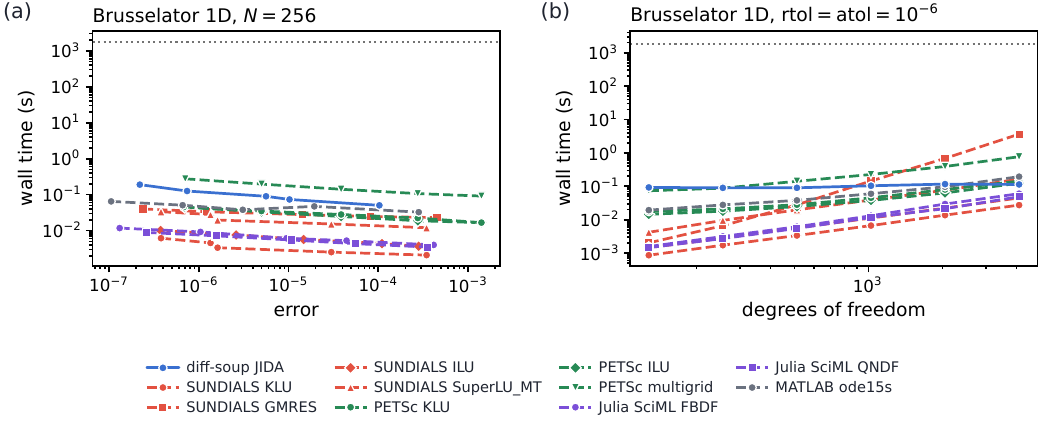}
\caption{
Work-precision (a) and scaling (b) diagrams for the one-dimensional Brusselator for \texttt{diff-soup} (solid blue), SUNDIALS (dashed red), PETSc (dashed green), Julia \texttt{DifferentialEquations.jl} (dashed purple), and MATLAB (dashed gray), with the marker denoting the particular solver as described in the legend.
All benchmarks are run according to the procedure described in Section~\ref{subsec:benchmarks}.
Notably, data points are omitted where the solve did not converge or exceeded the wall cap.
Generally, we find that for a one-dimensional problem below the crossover of Figure~\ref{fig:heat-1d}d the fixed overhead of the GPU dominates and the CPU based solvers lead.
(a) The work-precision diagram shows the CPU based solvers outperforming \texttt{diff-soup}, the fastest of them by over an order of magnitude, with only the multigrid-preconditioned PETSc solver trailing it.
(b) The scaling diagram shows \texttt{diff-soup} exhibiting a fixed overhead across the ladder, whereas the CPU based solvers exhibit power-law growth.
}
\label{fig:brusselator-1d-forward}
\end{figure}

\bibliographystyle{elsarticle-num-names}
\bibliography{references.bib}

\end{document}